\RequirePackage{lineno} 
\documentclass[preprint]{aastex}

\usepackage{natbib}
\usepackage{times}
\usepackage{hyperref}
\usepackage{longtable}
\usepackage{afterpage}

\slugcomment{to appear in Astronomical Journal
(submitted 09 August 2016)}

\newcommand{\mpg}{{g'}}
\newcommand{\mpr}{{r'}}

\newcommand{\NKBO}{21\ }
\newcommand{\lKBO}{24\ }
\newcommand{\TKBO}{21\ }

\newcommand{\IKBO}{7\ }

\shorttitle{HiLat Data Release}
\shortauthors{Petit {\em et al.}}

\setlongtables

\begin{document}
\linenumbers

\title{The Canada-France Ecliptic Plane Survey (CFEPS) - High Latitude
  Component}

\author{J-M. Petit\altaffilmark{a,b}, 
J.J. Kavelaars\altaffilmark{c},       
B.J. Gladman\altaffilmark{b},         
R.L. Jones\altaffilmark{b,c},         %
J.Wm. Parker\altaffilmark{d},         %
C. Van Laerhoven\altaffilmark{e, {\rm \ now\ at\ } i}, 
R. Pike\altaffilmark{f},              
P. Nicholson\altaffilmark{g},         %
A. Bieryla\altaffilmark{d, {\rm \ now\ at\ } h},           %
M.L.N. Ashby\altaffilmark{h},         
S.M. Lawler\altaffilmark{b, {\rm \ now\ at\ } c} 
}

\altaffiltext{a}{Institut UTINAM, CNRS-UMR 6213, Observatoire de Besan\c{c}on,
  BP 1615, 25010 Besan\c{c}on Cedex, France}
\altaffiltext{b}{Department of Physics and Astronomy, 6224
  Agricultural Road, University of British Columbia, Vancouver, BC, Canada}
\altaffiltext{c}{Herzberg Institute of Astrophysics, National Research
  Council of Canada, Victoria, BC V9E 2E7, Canada}
\altaffiltext{d}{Planetary Science Directorate, Southwest Research
  Institute, 1050 Walnut Street, Suite 300, Boulder, CO 80302, USA}
\altaffiltext{e}{Department of Planetary Sciences, University of Arizona, 1629
  E. University Blvd, Tucson, AZ, 85721-0092, USA}
\altaffiltext{f}{Department of Physics and Astronomy, University of Victoria,
  Victoria, BC, Canada}
\altaffiltext{g}{Cornell University, Space Sciences Building, Ithaca, New York
  14853, USA}
\altaffiltext{h}{Harvard-Smithsonian Center for Astrophysics, 60 Garden
  Street, Cambridge, MA 02138, USA}
\altaffiltext{i}{Canadian Institute for Theoretical Astrophysics, 60 St George
  St., Toronto, ON, M5S 3H8, Canada}

\begin{abstract}
We report the orbital distribution of the Trans-Neptunian objects (TNOs)
discovered during the High Ecliptic Latitude (HiLat) extension of the
Canada-France Ecliptic Plane Survey (CFEPS), conducted from June 2006 to July
2009.  The HiLat component was designed to address one of the shortcomings of
ecliptic surveys (like CFEPS), their lack of sensitivity to high-inclination
objects.  We searched 701~deg$^2$ of sky ranging from 12$^\circ$ to 85$^\circ$
ecliptic latitude and discovered \lKBO TNOs, with inclinations between
15$^\circ$ to 104$^\circ$. This survey places a very strong constraint on the
inclination distribution of the hot component of the classical Kuiper Belt,
ruling out any possibility of a large intrinsic fraction of highly inclined
orbits. Using the parameterization of \citet{2001AJ....121.2804B}, the HiLat
sample combined with CFEPS imposes a width $14^\circ \le \sigma \le
15.5^\circ$, with a best match for $\sigma = 14.5^\circ$.  HiLat discovered the
first retrograde TNO, 2008~KV$_{42}$, with an almost polar orbit with
inclination 104$^\circ$, and (418993),
a scattering object with
perihelion in the region of Saturn's influence, with $a \sim 400$~AU and $i =
68^\circ$.

\end{abstract}

\keywords{Kuiper Belt, surveys}

\pagebreak

\section{Introduction}

The Kuiper Belt is widely thought of as a left-over flattened disk of
planetesimals extending from $\sim30$ to a thousand AU from the Sun. Several
Kuiper Belt surveys broke ground by investigating the gross properties of the
TNO diameter and orbital distributions via large samples
\citep{1996AJ....112.1225J,2001AJ....122.1051G,2002AJ....123.2083M,2001AJ....122..457T}. It
is now obvious that this region must have been heavily perturbed late in the
process of giant planet formation.  The Kuiper Belt's small mass and the
existence of many objects with large orbital inclinations ($i$ up to
50$^{\circ}$) indicate that a process either emptied most of the mass out of
the primordial Kuiper Belt or, more dramatically, that the Kuiper Belt was
transported to its current location during planetary migration. Recent models
suggest stellar encounters (e.g.,
\citet{2010Sci...329..187L,2012Icar..217....1B}) or the existence of a 9th
planet \citep{2016AJ....151...22B} may play an important role in shaping the
outer solar system.

The dynamical structure of the Kuiper Belt is much more complex than
anticipated by the community. Surveys with known high-precision detection
efficiencies and which track essentially all their objects, to avoid ephemeris
bias \citep{2008ssbn.book...59K,2010AJ....139.2249J}, are needed to disentangle
these details and the cosmogonic information they provide. The Canada-France
ecliptic plane survey (CFEPS)\footnote{{\it http://www.cfeps.net}} \citep[][P1
  hereafter]{2006Icar..185..508J,2009AJ....137.4917K,2011AJ....142..131P}, was
a fully characterized\footnote{A survey is characterized when all detection
  circumstances are known: telescope pointings, efficiency of detection
  w.r.t. magnitude and apparent motion, ..., so that one can simulate the
  survey. It is fully characterized if tracking has no orbital bias. An object
  is characterized when its detection efficiency is large enough that it is
  accurately determined \citep{2004MNRAS.347..471P}} survey that tracked more
than 80\% of its discoveries to orbit classification\footnote{Assigning an
  orbit to a dynamical class, as defined by
  \citet{2008ssbn.book...43G}}. Although discovering and tracking only 169
TNOs, this survey produced solid science contributions to Kuiper Belt science
\citep[P1;][]{2006Icar..185..508J,2009AJ....137.4917K,2012AJ....144...23G}. Without
this accurate calibration and extensive tracking, it is risky to perform
quantitative interpretation of the orbital distribution of the ∼800
multi-opposition TNOs in MPC database with unknown detection and tracking
biases \citep{2010AJ....139.2249J}.

The inclination distribution of the `main' Kuiper Belt is now recognized as
bimodal \citep{2001AJ....121.2804B,2008ssbn.book...59K}, with a `cold'
component of objects with inclination width around 3$^\circ$ and a `hot'
component with a very broad inclination distribution, much like the disk/halo
structure of the galaxy.  This discovery came at the same time as the
realization that the cold component appears to have a different colour
distribution than the hot component
\citep{2002AJ....124.2279D,2003ApJ...599L..49T,2012ApJ...749...33F,2015A&A...577A..35P}. The
orbital distribution of these high-inclination objects has a huge lever arm on
models of outer Solar System formation and evolution, which include ideas like
passing stars
\citep{2000ApJ...528..351I,2004Natur.432..598K,2004AJ....128.2564M,2011Icar..215..491K}
that predict mean inclinations increasing with semimajor axis, rogue planets
\citep{2006ApJ...643L.135G} that predict inclination decreasing with semimajor
axis or transplanting almost all TNOs to their current locations during a
large-scale reorganization of the planetary system
\citep{1999Natur.402..635T,2008Icar..196..258L,2015AJ....150...73N}.

For both components the distribution of orbital inclination can be modelled as
$P(i) \propto \sin{(i)} \exp{(-i^2/2\sigma^2)}$
\citep{2001AJ....121.2804B}. The distribution of the hot component appears to
have a Gaussian width $\sigma$ of at least 15$^\circ$
\citep[P1;][]{2001AJ....121.2804B,2009AJ....137.4917K,2010AJ....140..350G}, but
constraining the largest inclinations is difficult because detection biases in
ecliptic surveys strongly disfavour their discovery. About two dozen TNOs with
orbital inclinations in excess of 40$^\circ$ are now known. Eris, the belt's
most massive known member \citep{2005ApJ...635L..97B}, is in this group along
with 2004~XR$_{190}$ \citep[discovered by our group during
  CFEPS;][]{2006ApJ...640L..83A}, the lowest-$e$ orbit known TNO with
semi-major axis beyond 50AU.

\begin{figure*}[H]
\begin{center}
\includegraphics[width=15cm]{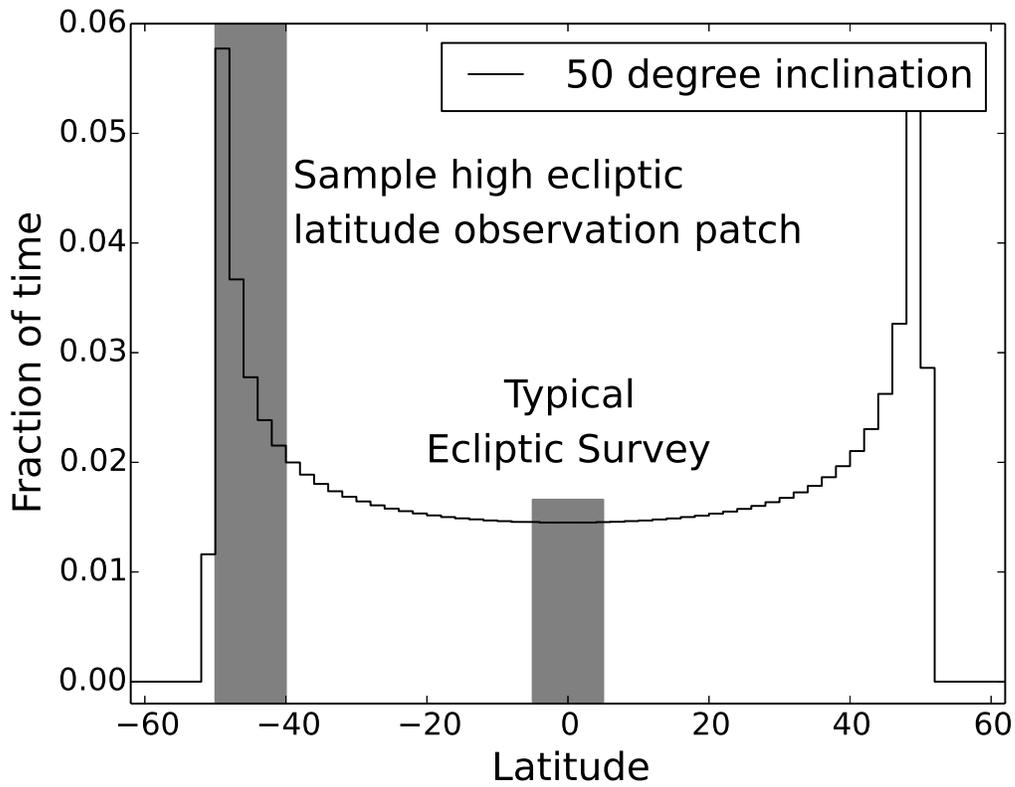}
\end{center}
\caption{Fraction of time spent at each ecliptic latitude for a sample object
  with an orbital inclination of 50 degrees. Previous surveys have mostly
  concentrated on low ecliptic latitudes where their sensitivity to high
  inclinations objects is comparatively low (central grey region). A survey
  concentrating on the area between 40--50$^\circ$ latitude (like parts of
  HiLat, see Table~\ref{tab:fields} and Fig.~\ref{fig:hilat_geom}, left grey
  region), where high inclination objects spend much of their orbital period,
  would be more sensitive to these objects.  }
\label{fig:inc_time}
\end{figure*}

Kuiper Belt objects with large inclinations spend the majority of their time at
high ecliptic latitudes (Fig.~\ref{fig:inc_time}) and are poorly represented in
the ecliptic surveys (including the main component of CFEPS). Even more
dramatically, it has become clear that the size distribution of the high
inclination component is flatter (number of objects increases slower when size
decreases) than the ecliptic component
\citep[P1;][]{2001AJ....121.1730L,2004AJ....128.1364B,2014ApJ...782..100F}. So
deeper surveys concentrating on the ecliptic will be increasingly dominated by
low inclination objects.

\begin{figure*}[H]
\begin{center}
\includegraphics[width=15cm]{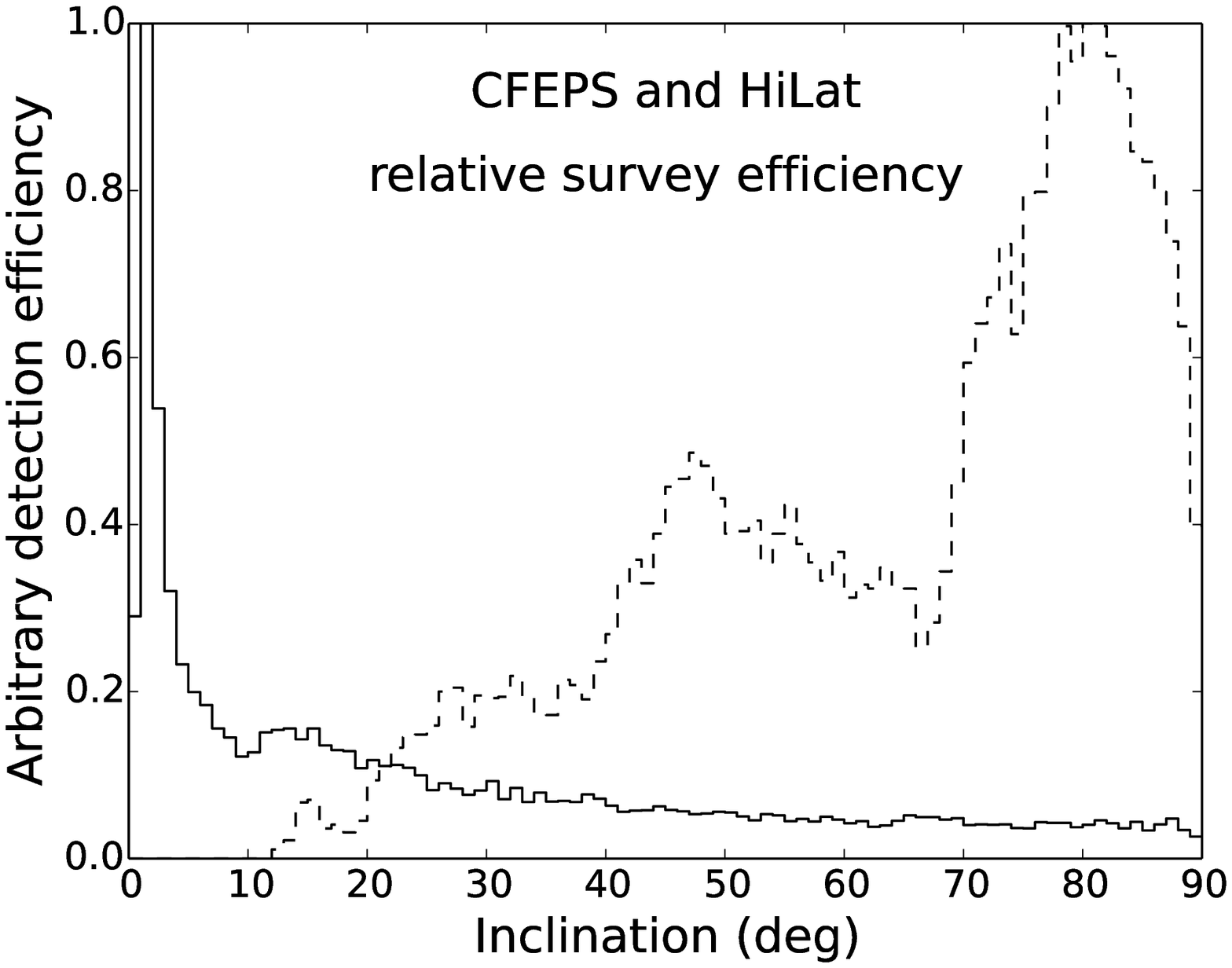}
\end{center}
\caption{An illustration of the contrasting detection efficiencies of CFEPS
  (solid line) and HiLat (dashed line) as a function of ecliptic inclination,
  given their actual pointing histories. The orbital distribution model used
  here is the one derived from CFEPS, except for the inclination which was
  drawn uniformly between 0--90$^\circ$. The scaling of each histogram is
  arbitrary, what matters here is the relative efficiency of a given survey to
  different inclinations.}
\label{fig:both_sensitivity}
\end{figure*}

The situation at the end of 2006 was that a large fraction of the sky within a
few degrees of the ecliptic had been covered by a few large surveys, with
magnitude limits in the range of $m_R$=20--23. Being insensitive to high
inclination objects (Fig.~\ref{fig:both_sensitivity}), ecliptic surveys have
poor sensitivity to the width of the hot population. Thanks to two deep blocks
of 11~deg$^2$ (one at 10$^\circ$ and another at 20$^\circ$ ecliptic latitude)
the CFEPS efficiency decreases less than most other ecliptic surveys towards
higher ecliptic latitudes. Still, although CFEPS preferes a hot population
inclination width $\sigma$ of 16$^\circ$, it could not reject a width of
25$^\circ$. Actually what limits the value of $\sigma$ is the relative decrease
of the number of low and intermediate inclination objects when increasing
$\sigma$. Using the converted Palomar Schmidt, \citet{2003EM&P...92...99T} had
examined the majority of the northern sky to a depth of $m_R\sim20.5$ (limit
for median observing conditions), discovering several of the largest known
objects; several of these large-inclination objects (like Eris) were close to
the depth and motion limits of that survey due to their great distances. The
ESSENCE Supernova Survey \citep{2008ApJ...682L..53B} announced the detection of
14 TNOs found in images covering $\sim$11~deg$^2$ to $\mpr\sim 23.7$ in the
ecliptic latitude range -21$^\circ$ to -5$^\circ$; this work also showed that
once outside of the ecliptic core, the sky density is consistent with even a
uniform distribution in latitude. Such a distribution would not be rejected by
any characterized surveys known at the time.  We decided to perform a deep
survey to magnitude $m_R\sim$~23.5--24.0 at high ($> 15^\circ$) ecliptic
latitudes, called HiLat, to probe the hot component of the Kuiper Belt at sizes
smaller than achieved by the Palomar wide area survey
\citep{2003EM&P...92...99T} and SDSS.  Although HiLat is insensitive to objects
with inclinations below 10$^\circ$ ecliptic latitude
(Fig.~\ref{fig:both_sensitivity}), it complements existing surveys because its
design makes it very sensitive to objects having inclinations beyond
20$^\circ$--30$^\circ$ (Fig.~\ref{fig:both_sensitivity}).

\afterpage{\clearpage}

This manuscript describes the observations carried out during the six years of
the project and provides our complete catalog (the HiLat release) of
off-ecliptic detections and characterizations along with fully linked
high-quality orbits.  In summary, the `products' of the HiLat survey consist
of four items:
\begin{enumerate}
\item A list of detected HiLat TNOs, associated with the sky location of
  discovery,
\item a characterization of each survey discovery observation (detection
  efficiency as a function of magnitude, motin on sky; rate range searched;
  pointing of observations; etc.),
\item a Survey Simulator that takes a proposed Kuiper Belt model, exposes
  it to the known detection biases of the HiLat blocks and produces simulated
  detections to be compared with the real detections, and
\item the updated CFEPS model populations accounting for the HiLat detections.
\end{enumerate}

\section{Observations and Initial Reductions}
\label{sec:obs}

\begin{figure*}[H]
\begin{center}
\includegraphics[width=16cm]{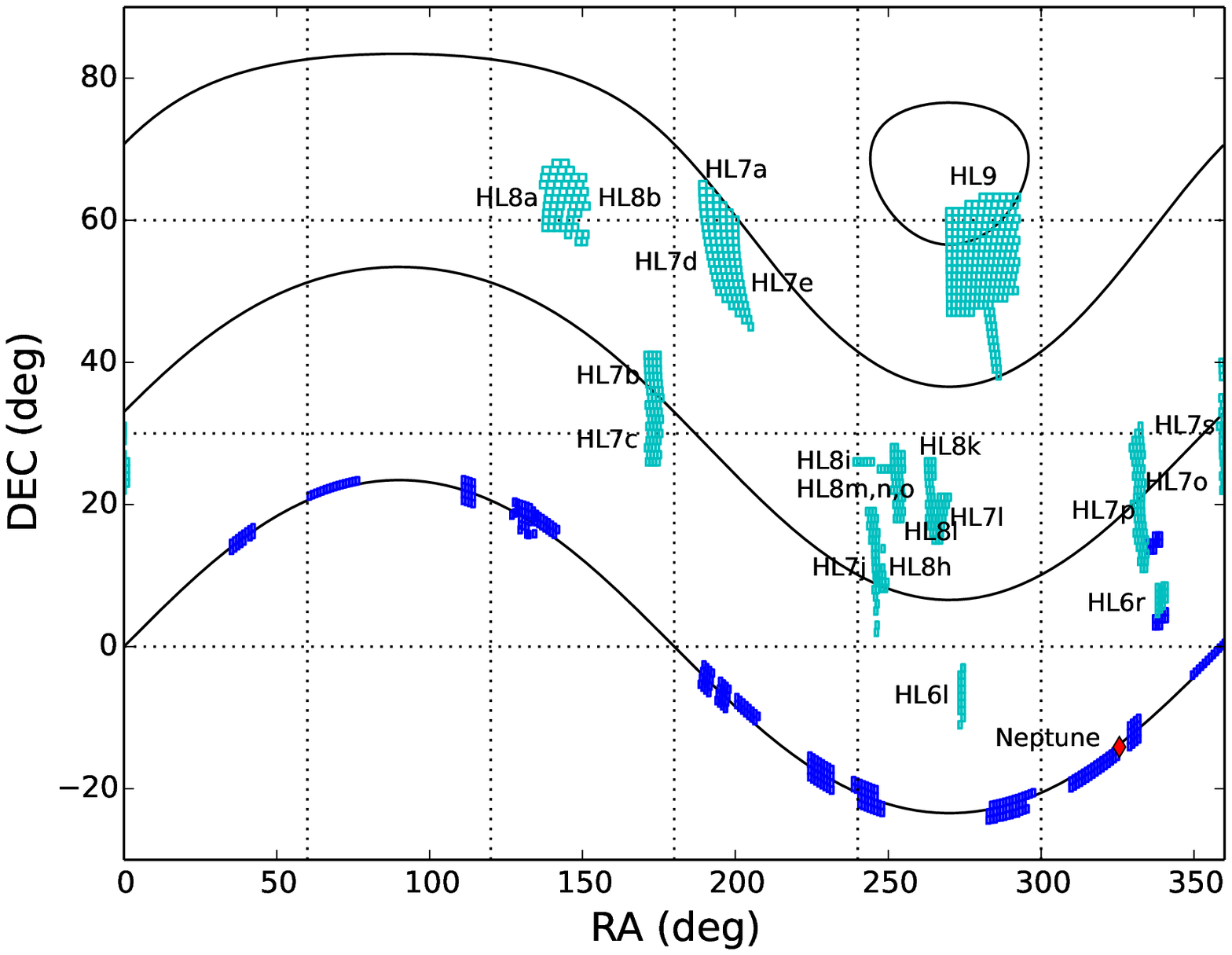}
\end{center}
\caption{Geometry of the HiLat discovery-blocks. The RA and Dec grid is
  indicated with dotted lines. The solid curves show constant ecliptic
  latitudes of 0$^\circ$, 30$^\circ$, 60$^\circ$, 80$^\circ$, from bottom to
  top. The blue rectangles mostly along the eclitpic indicate CFEPS pointings,
  the cyan rectangles indicate the HiLat survey pointings. The red diamond
  indicates the position of Neptune on 2016-07-31.  }
\label{fig:hilat_geom}
\end{figure*}

The discovery component of the HiLat project imaged $\sim$700 square degrees of
sky, all of which was at ecliptic latitude larger than 12$^\circ$, extending
almost to the North ecliptic pole (85$^\circ$, Fig.~\ref{fig:hilat_geom}).
Discovery observations, comprising a triplet of images 1 hour apart each on the
date listed in Table~\ref{tab:fields}, and a {\it nailing} observation, a
single image acquired a few nights away from the discovery triplet, were all
acquired using the Canada-France-Hawaii Telescope (CFHT) MegaPrime camera which
delivered discovery image quality (FWHM) of 0.7--0.9 arc-seconds in queue-mode
operations. The observations occured in {\it blocks} of 11 to 32 contiguous
fields, cycling three times between the fields. The number of fields observed
in a series was chosen such as to have $\sim$1~hour between two consecutive
observations of the same field. When a block was too large to be observed
within one night, it was split into two sub-blocks observed during close-by
nights, with similar observing conditions.
All discovery imaging data is publicly available from the Canadian Astronomy
Data Centre (CADC\footnote{http://www.cadc.hia.nrc.gc.ca}).

\begin{deluxetable}{lrrrrrrrlrccc}
\tabletypesize{\scriptsize}
\tablecolumns{12}
\tablewidth{0pt}
\tablecaption{Summary of Field positions and Detections.\label{tab:fields}}
\tablehead{
\colhead{Block}  &  \colhead{RA} &   \colhead{Dec}   &  \colhead{Area}  &
\colhead{Fill} & \multicolumn{2}{c}{Detections} &
\colhead{Ecl. lat.} & \multicolumn{2}{c}{Discovery} & \colhead{limit} &
\multicolumn{2}{c}{Detection limits}\\ 
\colhead{\ } & \colhead{HRS} & \colhead{deg} &  \colhead{deg$^2$}  &
\colhead{Factor} & \colhead{D} & \colhead{T}  & \colhead{range (deg)} &
\colhead{date} & \colhead{filter} & \colhead{$r_{AB}$} &
\colhead{rate (``/h)} & \colhead{direction (deg)}
}
\startdata
HL6l   & 18:16  &  -06:49  &  15 &  0.80 & 0 & 0 & 11:50--20:50 & 2006-06-23  &  r.MP9601 & 22.37 &    0.5 to   6.1 &  -17.8 to  16.4 \\
HL6r   & 22:37  &  +07:04  &  16 &  0.80 & 7 & 6 & 12:20--16:40 & 2006-09-18  &  r.MP9601 & 23.89 &    0.5 to   6.1 &  -43.6 to  -8.2 \\
\\
HL7a   & 13:06  &  +55:00  &  32 &  0.90 & 0 & 0 & 49:40--60:00 & 2007-03-18  &  r.MP9601 & 23.58 &    0.5 to   5.7 &    6.6 to  47.8 \\
HL7b   & 11:33  &  +37:30  &  32 &  0.88 & 0 & 0 & 27:00--35:40 & 2007-03-23  &  r.MP9601 & 22.89 &    0.5 to   5.4 &  -10.0 to  33.8 \\
HL7c   & 11:33  &  +29:30  &  32 &  0.89 & 4 & 4 & 19:50--28:40 & 2007-03-21  &  r.MP9601 & 23.72 &    0.5 to   5.8 &   -3.1 to  36.3 \\
HL7d   & 12:49  &  +57:00  &  32 &  0.84 & 0 & 0 & 49:30--59:50 & 2007-04-09  &  r.MP9601 & 23.28 &    0.5 to   4.6 &  -25.7 to  26.9 \\
HL7e   & 13:23  &  +52:58  &  32 &  0.87 & 0 & 0 & 49:40--60:10 & 2007-04-22  &  r.MP9601 & 23.47 &    0.5 to   4.7 &  -25.2 to  27.0 \\
HL7j   & 16:22  &  +12:53  &  32 &  0.90 & 5 & 5 & 22:50--40:20 & 2007-06-12  &  r.MP9601 & 23.49 &    0.5 to   5.6 &  -22.0 to  19.8 \\
HL7l   & 17:47  &  +18:03  &  27 &  0.90 & 0 & 0 & 37:50--45:00 & 2007-06-12  &  r.MP9601 & 23.35 &    0.5 to   6.2 &  -10.1 to  22.1 \\
HL7o   & 22:12  &  +22:02  &  32 &  0.90 & 0 & 0 & 20:30--39:40 & 2007-08-20  &  r.MP9601 & 22.74 &    0.5 to   6.3 &  -28.0 to   1.2 \\
HL7p   & 22:06  &  +19:23  &  32 &  0.84 & 4 & 4 & 19:40--39:10 & 2007-09-06  &  r.MP9601 & 23.85 &    0.5 to   6.2 &  -41.3 to  -7.3 \\
HL7s   & 23:59  &  +27:54  &  31 &  0.98 & 0 & 0 & 19:30--37:00 & 2007-09-19  &  r.MP9601 & 23.38 &    0.5 to   6.3 &  -35.3 to  -3.9 \\
\\
HL8a   & 09:24  &  +63:30  &  30 &  0.90 & 1 & 1 & 40:00--50:20 & 2008-01-08  &  r.MP9601 & 23.76 &    0.6 to   6.6 &   22.1 to  50.1 \\
HL8b   & 09:52  &  +61:60  &  25 &  0.90 & 0 & 0 & 40:30--49:50 & 2008-01-09  &  r.MP9601 & 23.24 &    0.6 to   6.6 &   26.0 to  54.0 \\
HL8h   & 16:32  &  +09:58  &  11 &  0.88 & 0 & 0 & 29:10--35:50 & 2008-05-05  &  r.MP9601 & 23.91 &    0.5 to   6.2 &    5.4 to  35.8 \\
HL8i   & 16:21  &  +25:33  &  11 &  0.90 & 0 & 0 & 44:40--47:30 & 2008-05-09  &  r.MP9601 & 24.31 &    0.5 to   6.3 &    8.1 to  37.5 \\
HL8k   & 17:35  &  +24:25  &  12 &  0.90 & 1 & 1 & 44:50--49:50 & 2008-05-11  &  r.MP9601 & 24.63 &    0.5 to   6.4 &   16.4 to  41.0 \\
HL8l   & 17:36  &  +19:15  &  13 &  0.90 & 0 & 0 & 39:40--45:50 & 2008-05-13  &  r.MP9601 & 24.15 &    0.5 to   6.3 &   13.0 to  38.8 \\
HL8m   & 16:58  &  +23:15  &  12 &  0.90 & 0 & 0 & 39:50--49:50 & 2008-05-30  &  r.MP9601 & 24.26 &    0.5 to   6.1 &   -7.9 to  26.1 \\
HL8n   & 16:53  &  +22:33  &  11 &  0.89 & 1 & 1 & 39:40--50:30 & 2008-05-31  &  r.MP9601 & 24.80 &    0.5 to   6.1 &   -9.1 to  25.5 \\
HL8o   & 16:48  &  +23:00  &  12 &  0.90 & 0 & 0 & 39:30--50:20 & 2008-06-07  &  r.MP9601 & 24.26 &    0.5 to   5.8 &  -17.7 to  21.1 \\
\\
HL9    & 18:45  &  +55:08  & 219 &  0.92 & 1 & 1 & 59:30--85:20 & 2009-06-16  &  r.MP9601 & 24.28 &    0.5 to  20.0 &  -20.0 to 90.0 \\
\\
& \multicolumn{2}{r}{Grand Total} & 701 &  &  \lKBO & \TKBO  \\
\\
\enddata
\tablecomments{RA/Dec is the approximate center of the
  field. Fill Factor is the fraction of the rectangle Area
  covered by the mosaic and useful for TNO searching.
  D is the number of TNOs detected in the block, T is the number of them that
  have been tracked to dynamical classification. Only one HL6r detection with
  apparent magnitude beyond the characterization limit, was not tracked to a
  high-quality orbit.
  The limiting magnitude of the survey, $r_{AB}$, is in the SDSS
  photometric system and corresponding to a 40\% efficiency of detection. 
  Detection limits give the limits on the sky motion in rate (``/hr) and
  direction (``zero degrees'' is due West, and positive to the North).
}
\end{deluxetable}

The HiLat designation of a block was: a leading `HL'
followed by the year of observations (6 to 9) and then a letter
representing the two week period of the year in which the search observations
were acquired (example: HL7j occured in the second half of May 2007), similar
to CFEPS naming scheme.  Discovery observations occurred between June 2006 and
June 2008 for the coverage below 60$^\circ$ ecliptic latitude, followed by
observations between 60$^\circ$ and 85$^\circ$ ecliptic latitude from May to
July 2009. This last part of the survey is simply named HL9 as it was acquired
as 22 contiguous blocks over this time span.

The discovery fields were chosen in order to maximize our sensitivity to the
latitude distribution of the Kuiper Belt, in particular the high inclination
TNOs. Observing at high ecliptic latitude ensured that we observed only
high-inclination TNOs, and greatly decreased the pressure for follow-up
observations, as the number of TNOs per unit area drops sharply away from the
ecliptic. The ecliptic longitudes were chosen to avoid the galactic plane, and
maximize our chances to get discovery and tracking observations (due to typical
weather at time of opposition for the discovery field, observing request
pressure on the telescope). Each of the discovery blocks was searched for TNOs
using our Moving Object Pipeline \citep[MOP; see][]{2004MNRAS.347..471P}.
Table~\ref{tab:fields} provides a summary of the survey fields, imaging
circumstances and detection thresholds.  Subsequent tracking, over the next 2
or more oppositions, occurred at a variety of facilities, including CFHT,
summarized in Table~\ref{tab:followup}.  The field sequencing and follow-up
strategy of this survey are similar to those of CFEPS
\citep{2006ApJ...640L..83A,2009AJ....137.4917K,2011AJ....142..131P}.  Our
discovery and tracking observations were made using short exposures designed to
maximize the efficiency of detection and tracking of the TNOs in the
field. These observations do not provide the high-precision flux measurements
necessary for possible taxonomic classification based on broadband colours of
TNOs and we do not comment here on this aspect of the HiLat sample.

\begin{deluxetable}{llrc|cllr}
\tablecolumns{7}
\tablewidth{0pc}
\tablecaption{Follow-up/Tracking Observations.\label{tab:followup}}
\tablehead{
\colhead{UT Date}    & \colhead{Telescope} & \colhead{Obs.} &
\colhead{\ \ } & \colhead{\ \ } & \colhead{UT Date}    & \colhead{Telescope} &
\colhead{Obs.}} 
\startdata

2006 Nov 22 & WIYN 3.5-m         &   8 & & & 2008 Aug 31 & CFHT 3.5-m         &   6 \\ 
2007 Apr 13 & CFHT 3.5-m         &   6 & & & 2008 Oct 22 & WIYN 3.5-m         &   9 \\ 
2007 May 14 & Hale 5-m           &  13 & & & 2008 Dec 15 & Hale 5-m           &  13 \\ 
2007 May 14 & KPNO 2.1-m         &   7 & & & 2008 Dec 20 & WIYN 3.5-m         &  17 \\ 
2007 Jul 26 & CFHT 3.5-m         &   3 & & & 2009 Jan 26 & CFHT 3.5-m         &   7 \\ 
2007 Sep 10 & WIYN 3.5-m         &   8 & & & 2009 Mar 25 & Subaru 8.2-m       &   2 \\ 
2007 Sep 13 & CFHT 3.5-m         &  20 & & & 2009 Apr 22 & Subaru 8.2-m       &   5 \\ 
2007 Sep 15 & Hale 5-m           &  25 & & & 2009 Jun 19 & WIYN 3.5-m         &  30 \\ 
2007 Oct 07 & CFHT 3.5-m         &   6 & & & 2009 Jul 18 & CFHT 3.5-m         &   5 \\ 
2007 Nov 08 & WIYN 3.5-m         &  17 & & & 2009 Jul 23 & Hale 5-m           &  31 \\ 
2008 Mar 04 & CFHT 3.5-m         &  12 & & & 2009 Aug 17 & Hale 5-m           &   6 \\ 
2008 Mar 08 & CFHT 3.5-m         &   3 & & & 2009 Aug 18 & CFHT 3.5-m         &   6 \\ 
2008 Apr 04 & CFHT 3.5-m         &  10 & & & 2009 Sep 12 & CFHT 3.5-m         &   4 \\ 
2008 May 02 & WIYN 3.5-m         &  21 & & & 2009 Sep 13 & CFHT 3.5-m         &  27 \\ 
2008 May 05 & CFHT 3.5-m         &  21 & & & 2009 Oct 12 & CFHT 3.5-m         &   8 \\ 
2008 May 28 & CFHT 3.5-m         &  14 & & & 2009 Nov 15 & CFHT 3.5-m         &   4 \\ 
2008 Jun 01 & CFHT 3.5-m         &   3 & & & 2010 Jan 20 & CFHT 3.5-m         &   3 \\ 
2008 Jun 07 & CTIO 4-m           &  20 & & & 2010 Mar 19 & Hale 5-m           &  12 \\ 
2008 Jun 22 & MMT 6.5-m          &   4 & & & 2011 May 02 & Magellan 6.5-m     &   8 \\ 
2008 Jul 07 & Gemini South 8.1-m &   5 & & & 2013 Feb 06$^{(a)}$ & Gemini North 8.1-m &  42 \\ 
2008 Aug 05 & CFHT 3.5-m         &  24 & & & 2013 Jul 05 & NOT 2.5-m          &  13 \\ 
2008 Aug 30 & CFHT 3.5-m         &  52 & & & 2013 AUg 05$^{(a)}$ & Gemini North 8.1-m &  32 \\ 

\enddata \tablecomments{
  All observations not part of the HiLat discovery
  survey are reported here. UT Date is the start of the observing run;
  Obs. is the number of astrometric measures reported from the observing
  run.  Runs with low numbers of astrometric measures were either wiped out by
  poor weather, or not meant for HiLat object follow-up originally. $(a)$ This
  is the date of the first observation; targets were observed twice a month
  throughout the semester.}
\end{deluxetable}

\afterpage{\clearpage}

\section{Sample Characterization}
\label{sec:char}

As is now the norm
\citep{1998AJ....115.1680T,1998AJ....115.2125J,1998AJ....116.2042G,2000ApJ...529L.103T,2001AJ....122.1051G,2006MNRAS.365..429P,2009AJ....137.4917K,2011AJ....142..131P},
we characterized the magnitude-dependent detection probability of each
discovery block by inserting artificial sources in the images.
We performed differential aperture photometry for each of our detected
objects observed on photometric nights.
Our photometry is reported in the Sloan system \citep{1996AJ....111.1748F} with
the calibrations contained in the header of each image as provided by the
ELIXIR processing software \citep{2004PASP..116..449M}. It can be found in
Tables~\ref{tab:chclass} and \ref{tab:nonchclass}.
All HiLat discovery observations that detected TNOs were acquired in
photometric conditions in a relatively narrow range of seeing conditions due to
queue-mode acquisition.

Those real objects in each block that have a magnitude brighter than that
block's 40\% detection probability are considered to be part of the HiLat {\em
  characterized sample}.  Because detection efficiencies below $\sim$ 40\%
determined by human operators and our software diverge
\citep{2004MNRAS.347..471P}, and since characterization is critical to our
goals, we are unable to utilize the sample faint-ward of the measured 40\%
detection efficiency level for quantitative analysis (although we report these
discoveries, the majority of which were tracked to precise orbits).  The {\it
  characterized} HiLat sample consists of \NKBO\ objects of the
\lKBO\ discovered (Table~\ref{tab:chclass}).  The magnitude distribution of
objects detected brighter than our cutoff is consistent with the shape of the
TNO luminosity function \citep{2008ssbn.book...71P} and the typical decay in
detection efficiency due to gradually increasing stellar confusion and the
rapid fall-off at the SNR limit.

\begin{deluxetable}{ll|llrrrrrl}
\tabletypesize{\scriptsize}
\tablecolumns{7}
\tablewidth{0pc}
\tablecaption{Characterized Object Classification.\label{tab:chclass}}
\tablehead{
\multicolumn{2}{c|}{DESIGNATIONS} & \colhead{$a$} & \colhead{$e$} &
\colhead{$i$} & \colhead{$R$} &
\colhead{$m_r$} & \colhead{$\sigma_r$} & \colhead{$H_r$} & \colhead{Comment}\\
\colhead{CFEPS} & \multicolumn{1}{l|}{MPC} & \colhead{(AU)} & \colhead{\ }  &
\colhead{($^\circ$)} &  \colhead{(AU)}  & \colhead{\ }
& \colhead{\ } & \colhead{\ } & \colhead{\ }
}
\startdata

\cutinhead{Resonant Objects}

  HL6r3 & 2006 SG415 &  47.931(6) & 0.2915(1) & 31.376(0) & 35.009 & 23.27 & 0.03 & 7.75 & 2:1 \\ 
  HL7j3 & 2007 LG38  &  55.45(2) & 0.4340(3) & 32.579(0) & 32.219 & 22.93 & 0.09 & 7.68 & 5:2 \\ 
  HL7c1 & 2007 FN51  &  87.49(3) & 0.6188(2) & 23.237(0) & 39.100 & 23.20 & 0.06 & 7.17 & 5:1 I \\ 
  HL7j4 & 2007 LF38  &  87.57(3) & 0.5552(2) & 35.825(0) & 48.432 & 22.53 & 0.09 & 5.54 & 5:1 I \\ 

\cutinhead{Inner Classical Belt}  

  HL7p1 & 2007 RY326 &  38.817(9) & 0.06776(9) & 25.479(0) & 37.952 & 23.20 & 0.12 & 7.30 &  \\ 

\cutinhead{Main Classical Belt} 

  HL6r1 & 2007 RL314 &  40.386(8) & 0.0386(4) & 21.057(1) & 40.771 & 22.97 & 0.07 & 6.79 &  \\ 
  HL6r5 & 2006 SE415 &  42.599(8) & 0.027(1) & 18.517(1) & 42.266 & 23.73 & 0.12 & 7.40 &  \\ 
  HL6r6 & 2006 SF415 &  43.20(2) & 0.077(1) & 15.712(0) & 40.713 & 23.87 & 0.09 & 7.70 &  \\ 
  HL7c2 & 2007 FM51  &  45.53(1) & 0.159(1) & 29.221(1) & 42.561 & 23.00 & 0.15 & 6.59 &  \\ 
  HL7p2 & 2007 RW326 &  45.92(1) & 0.2355(2) & 20.500(0) & 35.127 & 23.70 & 0.10 & 8.16 & I (17:9) \\ 
  HL7p3 & 2007 RX326 &  46.096(7) & 0.1565(3) & 25.029(0) & 39.343 & 23.30 & 0.30 & 7.25 &  \\ 



\cutinhead{Detached/Outer Classical Belt}

  HL6r2 & 2006 SH415 &  49.759(9) & 0.2539(3) & 25.048(0) & 38.189 & 23.60 & 0.06 & 7.71 & \\ 
  HL7c3 & 2007 FO51  &  50.37(3) & 0.2873(6) & 27.946(0) & 37.560 & 22.87 & 0.19 & 6.99 & I (13:6) \\ 
  HL7j5 & 2007 LE38  &  54.05(1) & 0.2267(1) & 35.966(1) & 41.800 & 23.27 & 0.07 & 6.93 &  \\ 
  HL6r4 & 2007 RM314 &  70.81(2) & 0.4846(2) & 20.884(0) & 42.622 & 22.70 & 0.17 & 6.33 & I (18:5) \\ 
  HL7j1 & 2007 LJ38  &  72.37(3) & 0.4698(3) & 31.540(0) & 38.848 & 23.07 & 0.19 & 7.03 & I (15:4) \\ 
  HL8k1 & 2008 JO41  &  87.35(2) & 0.5431(1) & 48.815(0) & 44.453 & 24.57 & 0.12 & 7.91 &  \\ 

\cutinhead{Scattering Disk}

  HL8a1 & 2008 AU138 &  32.392(3) & 0.3745(2) & 42.826(1) & 44.518 & 22.93 & 0.23 & 6.29 &  \\ 
  HL8n1 & 2008 KV42  &  41.532(4) & 0.49138(7) & 103.447(0) & 31.849 & 23.73 & 0.03 & 8.52 &  \\ 
  HL7j2 & 2007 LH38  & 133.93(4) & 0.72523(8) & 34.197(0) & 37.376 & 23.37 & 0.03 & 7.50 & I (19:2) \\ 
  HL9m1 & 2009 MS9   & 348.9(2) & 0.96847(1) & 68.016(0) & 12.872 & 21.13 & 0.09 & 9.57 &  \\ 

\enddata
\tablecomments
{$a$: semimajor-axis (AU); $e$: eccentricity; $i$: inclination (degrees); $R$:
 distance to the Sun at discovery time (AU); 
 $m_r$: apparent magnitude of the object in MegaPrime $r'$ filter;
 $\sigma_r$: uncertainty on the magnitude in that filter;
 $H_r$ is the absolute magnitude in r band, given the distance at discovery;
 In Comment column, M:N: object in the M:N resonance;
 I: indicates that the orbit classification is insecure
 (see \citet{2008ssbn.book...43G} for an explanation of the exact meaning);
 (M:N): the insecure object may be in the M:N resonance.
 For the orbital elements the number in ``()'' gives the uncertainty on the
 last digit.}

\end{deluxetable}

\begin{deluxetable}{ll|llrrrrrl}
\tabletypesize{\scriptsize}
\tablecolumns{7}
\tablewidth{0pc}
\tablecaption{Non Characterized Object Classification.\label{tab:nonchclass}}
\tablehead{
\multicolumn{2}{c|}{DESIGNATIONS} & \colhead{$a$} & \colhead{$e$} &
\colhead{$i$} & \colhead{$R$} &
\colhead{$m_r$} & \colhead{$\sigma_r$} & \colhead{$H_r$} & \colhead{Comment}\\
\colhead{CFEPS} & \multicolumn{1}{l|}{MPC} & \colhead{AU} & \colhead{\ }  &
\colhead{$^\circ$} &  \colhead{AU}  & \colhead{\ }
& \colhead{\ } & \colhead{\ } & \colhead{\ }
}
\startdata

\cutinhead{Resonant Objects}

 uHL7c4 & 2007 FP51  &  44.760(6) & 0.2017(1) & 25.606(0) & 36.688 & 23.80 & 0.20 & 8.02  & 20:11 I \\ 

\cutinhead{Detached Classical Belt} 

 uHL7p4 & 2007 RZ326 &  52.676(8) & 0.3465(1) & 37.268(0) & 38.300 & 23.93 & 0.09 & 7.98 &  \\ 

\cutinhead{Non classified objects}

 uHL6r7 & 2006 SN415 & --- & --- & --- & 38(7) & 24.50 & 0.25 & 8.65 &  \\ 

\enddata
\tablecomments
    {Same as Table~\ref{tab:chclass} for non characterized objects.}

\end{deluxetable}

\afterpage{\clearpage}

\section{Tracking}
\label{sec:track}

For typical (i.e., low ecliptic latitude) surveys to depth $\mpr\sim$~23.5--24,
the observing load of tracking observations to secure the objects and determine
their orbits represents many times the time spent for discovery. In
such a case, a $\sim$700 square degree survey with fully tracked objects would
be prohibitive.  However, because HiLat covers very high ecliptic latitudes,
the number of object per square degree at our limiting magnitude goes down
dramatically beyond 30--35$^\circ$ and we detected only \lKBO objects (\NKBO
characterized). Hence the tracking observing load was much lower than for an
ecliptic survey and 

All of the \NKBO characterized and 2 of the 3 non-characterized objects were
followed for at least 3 oppositions. Objects that still had uncertain dynamical
classifications were then followed up to 7 oppositions, mostly for resonant or
near-resonant objects. The global release of the complete observing record for
all HiLat objects is available from the MPC \citep{2015MPEC....D....1P} and the
entire astrometric data for the HiLat objects can be found on the Besan\c{c}on
TNO database\footnote{{\it http://tnodb.obs-besancon.fr/}}. The correspondence
between HiLat internal designations and MPC designations can be determined
using Tables~\ref{tab:chclass} and \ref{tab:nonchclass} or from the
Besan\c{c}on TNO database. All characterized and tracked objects are prefixed
by {\it HL} and are used with the survey simulator for our modelling below.

The tracking observations provide sufficient information to allow reliable
orbits to be determined such that unambiguous dynamical classification can be
achieved in the majority of cases. Orbital elements are computed using the
\citet{2000AJ....120.3323B} `orbfit' code. Ephemerides errors for the coming
year are as small as a few tenths of an arcsecond for several objects, others
have uncertainties up to of order 10 arcseconds. Our protocol was to pursue
tracking observations until the semimajor axis uncertainty was $<0.1$\%; in
Tables~\ref{tab:chclass} and \ref{tab:nonchclass}, orbital elements are shown
to the precision with which they are known, with typical fractional accuracies
on the order of a few $10^{-4}$. In the cases of resonant objects even this
precision may not be enough to precisely determine the amplitude of the
resonant argument, or even securely classify them as resonant. Thanks to our
intensive tracking effort, dynamical classification is possible for 100\% of
the characterized sample.

\afterpage{\clearpage}

\subsection{Orbit classification}

We follow the dynamical classification scheme of \citet{2008ssbn.book...43G},
which was also used to determine the classification of the CFEPS sample.  In
this scheme, the Kuiper Belt is divided into three broad orbital classes based
on orbital elements and dynamical behavior. We first check if the object is
resonant (currently in MMR with Neptune or Uranus), then see if it is currently
scattering (practically defined as a variation of semimajor axis of more than
1.5~AU in a forward time integration over 10~Myr). If not, it is a classical or
detached object: Inner classical if semimajor axis is interior to the MMR 3:2
with Neptune; main classical if semimajor axis between the 3:2 and 2:1 MMR;
outer classical if semimajor axis beyond the 2:1 MMR and $e < 0.24$; detached
if semimajor axis beyond the 2:1 MMR and $e > 0.24$).

Using this classification procedure, \IKBO of our \NKBO characterized objects
remain insecure, as defined in \citet{2008ssbn.book...43G}, due to their
proximity to a (high-order) resonance border where the remaining astrometric
uncertainty makes it unclear if the object is actually resonant.  We list these
``insecure'' objects in the category shown by the majority of the clones
\citep{2008ssbn.book...43G} and give the nearby resonance in the comment
column. Table~\ref{tab:chclass} gives the classification of all characterized
objects. None of these objects had archival observations before our
discovery. Table~\ref{tab:nonchclass} gives the classification of the tracked
objects below the 40\% detection efficiency threshold, hence deemed
un-characterized and not used in our Survey Simulator comparisons.

The apparent motion of TNOs in our opposition discovery fields is approximately
$\theta$("/hr)~$\simeq$~(147~AU)/$R$, where $R$ is the heliocentric distance in
AU. With a typical seeing of 0.7--0.9 arcsecond and a time base of 70--90
minutes between first and third frames, we were sensitive to objects as distant
as $R \simeq$~125~AU, provided they are brighter than our magnitude
limit. Despite this sensitivity to large distances, the most distant object
discovered in HiLat lies at 48.4~AU from the Sun (HL7j4, an insecure resonant
object in the 5:1 MMR with Neptune \citep{2015AJ....149..202P}).

\section{Results}
\label{sec:results}

CFEPS data presented in P1 were modelled independently for the inner, main,
outer/detached classical, the scattering and various resonant populations by P1
and \citet{2012AJ....144...23G}. The model for the main classical belt is
refered as the L7 model hereafter. According to P1, the cold component may
very well exist only in the main classical belt. The hot component, on the
contrary, permeates the whole belt, from the inner classical, to the main
classical, to the outer/detached belt and all the resonances. The cold
component was well constrained by the Ecliptic component of the survey.

HiLat was designed to have maximum sensitivity to high-inclination objects
(Fig.~\ref{fig:both_sensitivity}), and thus places strong constraints on the
distribution of high-inclination objects, i.e., the hot population. The goal is
thus to improve the L7 model.

\afterpage{\clearpage}

\subsection{Main Classical belt and L7 model}

Our aim is to create a model that is compatible with both the CFEPS and HiLat
detections. We are able to account for HiLat detections by slightly changing
some parameters of the L7 orbital model, affecting only regions of phase space
not well constrainted by CFEPS detections.  Here we concentrate on the model
for the main classical belt, because this dynamical class alone constitutes
nearly a third of the full HiLat sample. With the parameterization of L7 model,
HiLat is sensitive almost exclusively to the hot component. Hence this is the
part of the model that will be modified in the following. However, in what
follows, we always run the full L7 model, including all components: kernel,
stirred and hot components.

\subsubsection{Orbital model}


To estimate the quality of a model, we compare the survey detected sample to
the sample returned by passing our intrinsic model through a survey simulator
\citep[see][for details]{2006Icar..185..508J}. Acceptance of a model is based
on the Anderson-Darling statistic for each of $a$, $e$, $i$, $q$ [perihelion
  distance], $R$ and $\mpr$ and its level of significance $s$ (probability of
the null hypothesis [the simulated and the observed samples are drawn from the
  same underlying distribution] being correct), determined using a bootstrap
method \citep{1992nrfa.book.....P}. $1 - s$ gives the rejectability of that
hypothesis. As for CFEPS, we reject a model when the rejectability exceeds
95\%. We determine the rejectability on the maximum of all 6 indicators we
consider. When creating the L7 model, P1 split the phase space into sub-regions
(see Appendix A of P1) to help separate the hot and cold components and account
for the kernel and stirred components. HiLat detects almost exclusively the hot
component, and the sample size is small, thus we determine the significance
examining the full orbital phase space occupied by the main-belt.

Using the improved survey simulator (see \citet{2016AJ.....inpressB} for a
description of the improvements) against the CFEPS detections, the L7 model for
the main classical belt retains the same level of significance ($\sim$20\%) as
with the previous survey simulator.

To combine the CFEPS and HiLat sample we must make a colour correction. CFEPS
was run mostly with the $\mpg$ filter, except for 1 block with the $\mpr$
filter, and the pre-survey block with the R filter. HiLat was run entirely with
the $\mpr$ filter. The improved Survey Simulator correctly handles surveys
observed in different filters, and accepts as input the colours of each
object. Here, for compatibility with previous works, we assume $\mpg - \mpr =
0.7$ and $\mpg - $R$ = 0.8$ \citep[this assumption agrees with more recent
  results from OSSOS, the Outer Solar System Origin
  Survey;][]{2016AJ.....inpressB}.

When the biased L7 model is tested agaijnst the HiLat detections, the $i$ and
$q$ distributions of the hot component are rejectable at $> 95$\%.  An
important feature of the L7 model for the main classical belt is the $q$
distribution of the hot component (see Appendix A of P1), which is essentially
uniform between two limits, with rapid roll-over at both ends, with a width of
0.5~AU. The upper limit is poorly constrained by CFEPS. To account for HiLat
detections, we moved the upper roll-over of the hot-component $q$ distribution
from 40 to 41~AU, still with a width of 0.5~AU. Because HiLat did not detect
any main classical belt object with $q < 35$~AU, we must impose a sharp cut-off
on top of the $i$-dependent lower-limit of the hot-component $q$ distribution.
The new parameterization is described in Appendix~\ref{sec:app_a}.  Using this
slight tuning of the L7 model continues to provide an acceptable match to the
CFEPS detected sample, when considered independent of the HiLat sample.
Extending the $q$-distribution of the L7 model somewhat allows compatibility
with the HiLat $q$-distribution.

\begin{figure*}[H]
\begin{center}
\includegraphics[width=15cm]{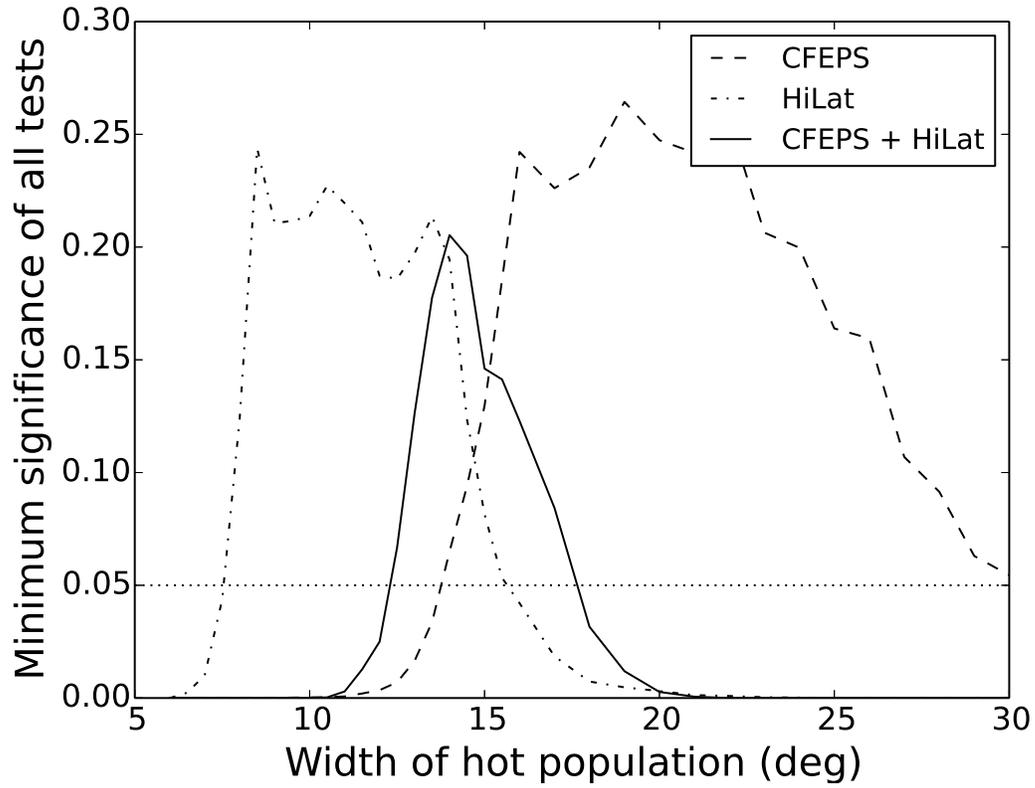}
\end{center}
\caption{Modelling the hot component of the main classical belt. Minimum level
  of significance of all 6 Anderson-Darling tests for $a$, $e$, $i$, $q$, $R$,
  and $\mpr$, as a function of the hot population inclination distribution
  width, for CFEPS alone (dashed line), HiLat alone (dash-dotted line) and
  CFEPS and HiLat combined (solid line). The dotted line shows the 95\%
  rejection threshold; any model with significance level below that line is
  rejected. The bumpiness of the curves is due to randomness in the survey
  simulator and in the bootstrapping of the Anderson-Darling statistics.}
\label{fig:minsigwidth}
\end{figure*}

The $i$-distribution of the HiLat main classical belt detected sample is
incompatible with the hot component of the L7 model. The CFEPS detected sample
strongly rejects a hot population with a narrow inclination width because that
model does not yield the correct ratio between low inclination and high
inclination as compared to the detections in the CFEPS sample. The CFEPS sample
rejects much larger inclination distributions ($\sigma \ge 30^\circ$; see
Fig.~\ref{fig:minsigwidth}, dashed line) only because of the relative lack of
low inclination objects in these distributions. The HiLat detected sample, on
the contrary, rejects any model with too wide an inclination distribution
because this survey is very sensitive to the high inclination orbits. Even the
inclination width $\sigma = 16^\circ$ preferred by CFEPS has a long tail
containing too many objects with $i > 35^\circ$ which would have been detected
by HiLat. But being completely insensitive to low inclination orbits (HiLat
cannot detect any of them), it can accept any values of $\sigma$ as long at
they allow enough objects up to $i \simeq 35^\circ$. Thus HiLat is consistent
with all values of $\sigma$ from 7.5$^\circ$ to 15.5$^\circ$
(Fig.~\ref{fig:minsigwidth}, dash-dotted line).  Together, the two surveys
combine high CFEPS sensitivity at low inclinations and HiLat's improved
sensitivity at high inclinations. The result is shown in
Fig.~\ref{fig:minsigwidth}. Because our model rejection threshold is set at 5\%
significance, this analysis indicates that an acceptable value for each of
CFEPS and HiLat separately and for their combination, is an inclination width
$\sigma$ in the range 14$^\circ$--15.5$^\circ$, where all three curves exceed
the threshold.

\afterpage{\clearpage}

Separately, CFEPS and HiLat favor different values for the width and only
marginally agree at the intersection (see Fig.~\ref{fig:minsigwidth}). There is
tension between the models allowed by the two data sets. This raises doubts on
the parameterization used here. \citet{2010AJ....140..350G} introduced an
inclination distribution given by $\sin{(i)}$ times a Gaussian of width
$\sigma$, centered on a value $i_c$ greater than 0$^\circ$ to fit what they
called the Scattered population
(Appendix~\ref{sec:app_a}). \citet{2015AJ....149..202P} did the same to study
the 5:1 MMR population. P1 mentioned the possibility to use a similar
functional form to represent the Classical belt hot population inclination
distribution, but concluded that the fit was good enough with the usual
distribution and that the data did not demand the increased complexity of the
extra parameter. Applying this functional form to the CFEPS, HiLat and
CFEPS+HiLat sample also does not improve the level of significance enough to
warrant the increased complexity of the extra parameter. So our preferred model
retains the previous inclination distribution functional form, with a width
$\sigma = 14.5^\circ$. We note, however, that the functional form here, while
useful for discussion, is not a good description of the physical distribution
of high-inclination TNOs.

\subsubsection{Population estimates}

Population estimates are dependent on the orbital model used to describe each
TNO component, which we are slightly modifying from P1.  They also depend on
the correct modelling of the survey operation and detection efficiency. As
explained in \citet{2016AJ.....inpressB}, the survey simulator has been
improved to better represent the exact selection and rejection effects of
objects based on measured magnitude rather than intrinsic magnitude. This has
the potential of substantially affecting the population estimates due to the
steep slopes of the absolute magnitude H distributions.

We follow the same procedure as in \citet{2009AJ....137.4917K},
\citet{2012AJ....144...23G}, and P1. We run our model, generating simulated
objects, passing them through the survey simulator until we have detected the
same number of objects in the simulation as in the real survey(s). We record
this number and repeat the procedure 500 times. This gives us the distribution
of likely population size.
Table~\ref{tab:pops}, columns A, gives the population estimates, using our new
model, to $H_g \le 8.0$ to compare with P1. Compared to P1, we use the new
$q$-distribution and an $i$-distribution with width $\sigma_h = 14.5^\circ$.
Our CFEPS estimates are statistically undistinguishable from P1 estimates.

\begin{deluxetable}{lcccccc}
\tabletypesize{\scriptsize}
\tablecolumns{5}
\tablewidth{0pt}
\tablecaption{Model dependent population estimates for $H_g \le 8.0$.\label{tab:pops}}
\tablehead{
\colhead{Population} & \multicolumn{2}{c}{CFEPS} & \multicolumn{2}{c}{HiLat} &
\multicolumn{2}{c}{CFEPS+HiLat}
 \\
\colhead{\ } &
\colhead{A} & \colhead{B} &
\colhead{A} & \colhead{B} &
\colhead{A} & \colhead{B}
}
\startdata
hot     & $3,700^{+800}_{-700}$ & $3,500^{+700}_{-700}$ &
$2,100^{+1900}_{-1300}$ & $2,700^{+3100}_{-1700}$ &
$3,500^{+700}_{-600}$ & $3,400^{+600}_{-600}$ \\
\\
stirred & $2,700^{+600}_{-500}$ & $2,600^{+500}_{-500}$ &
$1,550^{+1400}_{-950}$  & $2,000^{+2300}_{-1300}$ &
$2,600^{+500}_{-450}$ & $2500^{+450}_{-450}$ \\
\\
kernel  & $800^{+200}_{-150}$   & $750^{+150}_{-150}$ &
 $450^{+450}_{-300}$     & $600^{+700}_{-400}$ &
 $800^{+150}_{-150}$  & $750^{+150}_{-150}$  \\
\enddata
\tablecomments{Our model estimates are given for each sub-population within
the Kuiper belt. The uncertainties reflect 95\% confidence
intervals for the model-dependent population estimate.
Remember that the relative importance of each population
will vary with the upper $H_g$ limit. The A columns correspond to a uniform
colour $\mpg - \mpr$~=~0.7, while B columns have $\mpg - \mpr$~=~0.45 for the
hot component and $\mpg - \mpr$~=~0.95 for the cold component.
}
\end{deluxetable}

\afterpage{\clearpage}

Although the various population estimates for a given component have
overlapping error bars, HiLat estimates population sizes at just a little over
half those of CFEPS. This is also reflected in the larger than observed
fraction of objects detected from HiLat when running our model through the
combined CFEPS~+~HiLat survey simulator; 12\% of the simulated detections are
from HiLat, while they represent only 6\% of the real sample. This larger
fraction from HiLat means the model plus survey simulator are more efficient at
detecting objects in HiLat survey, hence needing a smaller underlying
population to reach the required number of detections. This may be due to our
choice of $\mpg - \mpr$ color for TNOs, a necessary parameter when combining
surveys done in different band passes.

Up to now we used the $\mpg - \mpr = 0.7$ colour derived from CFEPS sample for
all components. However, the cold belt objects are redder than the hot ones
\citep{2002AJ....124.2279D,2003ApJ...599L..49T}. If the hot objects detected by
HiLat are bluer than $\mpg - \mpr = 0.7$, then the number of objects brighter
than $H_g = 8.0$ needed to match the real detections is larger. According to
Fraser (private communication, 2016), the cold component has a typical colour
$0.8 < \mpg - \mpr < 1.1$, while the hot component comprise mostly neutral
objects with $0.4 < \mpg - \mpr < 0.7$, and a small fraction of objects as red
as the cold component. Table~\ref{tab:pops}, columns B, gives the population
estimates when using $\mpg - \mpr = 0.45$ for the hot component and $\mpg -
\mpr = 0.95$ for the cold component. The three population estimates become more
compatible with each other, and the fraction of simulated detections from HiLat
in CFEPS+HiLat simulations becomes 7\%, similar to the real detected
fraction. This result provides (unsurprising) evidence for the already known
different $\mpg - \mpr$ colours of the various components, which must be
accounted for when combining detections in different filters.

\subsection{Other populations}

The HiLat characterized sample included six outer classical or detached
objects, roughly half as many as were identified by CFEPS (P1 identified 13
non-scattering, non-resonant objects beyond 48~AU). P1 established that the
outer-detached population can be interpreted as a smooth extention beyond the
2:1 MMR of the hot main classical belt. We confirm this result with CFEPS+HiLat
detection. We note however that the HiLat sample alone allows inclination
widths $13^\circ < \sigma < 30^\circ$, possibly more excited than for the main
classical belt. The combined CFEPS+HiLat sample allows an inclination width
$12.5^\circ < \sigma < 20^\circ$. This is in agreement with the outer-detached
population being a smooth extension of the hot classical population. We
estimate the population beyond 48~AU $N(H_g \le 8.0) = 9500^{+4500}_{-3500}$,
very similar to P1 estimate.

The HiLat characterized sample contains 4 resonant objects. One is in the 2:1
MMR and another one in the 5:2 MMR with Neptune. These represent a small
contribution to the known populations of these resonances from characterized
surveys like CFEPS. HiLat made an important contribution to our understanding
of the resonant population by discovering two objects in the 5:1 MMR (only 1
was known from CFEPS), and another very close to the 5:1 MMR,
HL8k1~=~2008~JO$_{41}$ at 87.356~AU; scientific interpretation of these
discoveries have been reported in \citet{2015AJ....149..202P}.

\subsection{Exotic objects: 2008~KV$_{42}$ and (418993)~2009~MS$_9$}

Amongst its \NKBO characterized detections, HiLat discovered 2 extraordinary
TNOs. Both are scattering objects. The first one was discovered on May 31st,
2008 in a field at moderate ecliptic latitude ($\sim 30^\circ$). It is
HL8n1~=~2008~KV$_{42}$, the first known retrograde TNO. Details about this
object and what it tells us about the origin and dynamical evolution of exotic
scattering objects is developed in \citet{2009ApJ...697L..91G}.

\begin{figure*}[H]
\begin{center}
\includegraphics[width=15cm]{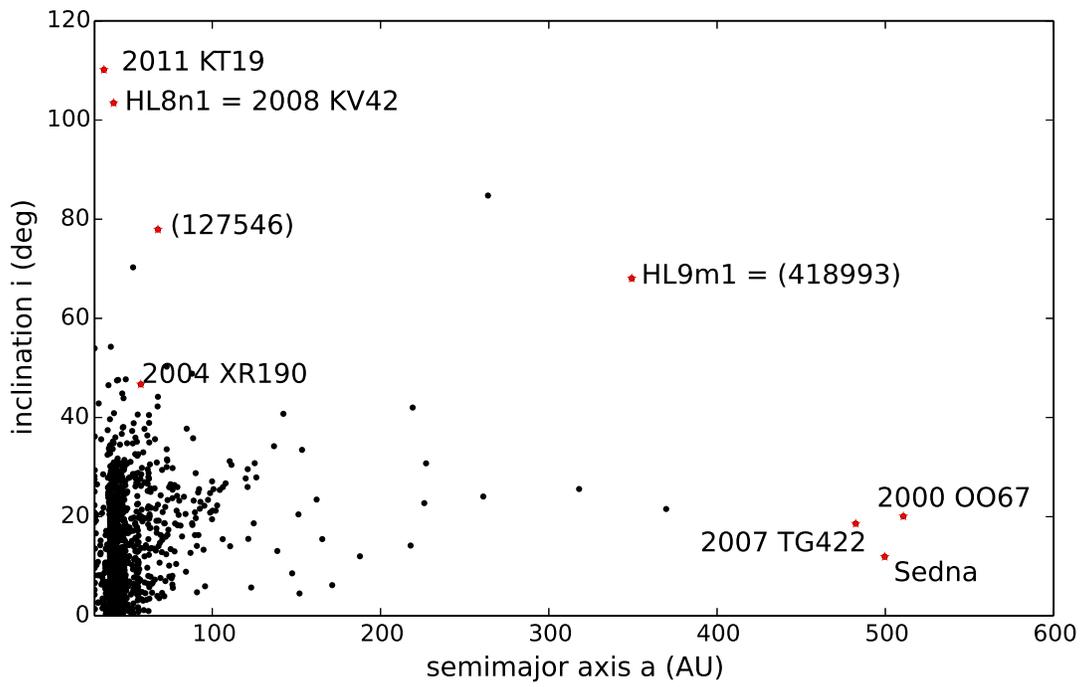}
\end{center}
\caption{Trans-Neptunian objects with $q > 10$~AU in orbital $a/i$ space, in
  ASTORB database as of August 2nd, 2016. Since its discovery,
  2009~MS$_9$ = (418993) stands out as unique (with other $a>300$~AU TNOs having
  inclinations in the `normal' $i<20$~deg range). 2008~KV$_{42}$ is also very
  peculiar with a retrograde orbit almost polar, having only one other object
  with similar orbit, 2011~KT$_{19}$. }
\label{fig:ms9-1}
\end{figure*}

The second object is HL9m1~=~(418993)~2009~MS$_9$, discovered on the 26th of
June 2009 at a distance of 12.9~AU from the Sun and an ecliptic latitude of
71$^\circ$. It has a large ($a \simeq 350$~AU) and highly-inclined ($i \sim
68^\circ$) orbit (Fig.~\ref{fig:ms9-1}), which is also highly eccentric ($e
\simeq 0.968$). Inbound at 13~AU at time of discovery, the pericenter of this
extreme orbit was $\sim$11~AU in February 2013, so (418993) is transiting
the range of heliocentric distances where comets have been observed to become
active \citep{2004come.book..317M}. (418993) thus may be the first
observable object that has been in deep cold storage at hundreds of AU for of
order 5,000 years. Under the hypothesis that this is a comet from a distant
source (either the inner Oort Cloud, or something else as yet unknown), it is
also quite possible that (418993) has never been interior to Saturn's orbit
(unlikely to be true for the known Centaurs, which often have their perihelia
altered as they interact with the giant planets).

\afterpage{\clearpage}

A plausible scenario is that (418993) is a former Oort-cloud object that has
had its orbit changed from nearly parabolic ($a>$1000~AU) to highly eccentric
by an encounter with Saturn, Uranus, or Neptune. (418993) is currently only
dynamically meta-stable on the order of $\sim$10~Myr, and may never have come
inside the water-sublimation zone (heliocentric distances of 5--6~AU). Many
comet nuclei have been studied after the development of a coma, but only after
the comets have left the inner Solar System and are very dim
\citep{2004cosp...35.1824L}. MS9 had the advantages that, at time of discovery,
it was bright ($\mpr\sim$ 22), inbound, and had no obscuring coma. Assuming an
albedo $p$=0.04 \citep[common for comet nuclei,][but on the lower end for
  TNOs]{2004cosp...35.1824L}, this object has a radius $\simeq$20~km. Not only
is (418993) unique dynamically, but if it had become an active comet, it
would have been the largest comet nucleus in recent times, after Hale-Bopp
\citep[C/1995 O1; radius = 37 km;][]{2004cosp...35.1824L}.

\begin{figure*}[H]
\begin{center}
\includegraphics[width=15cm]{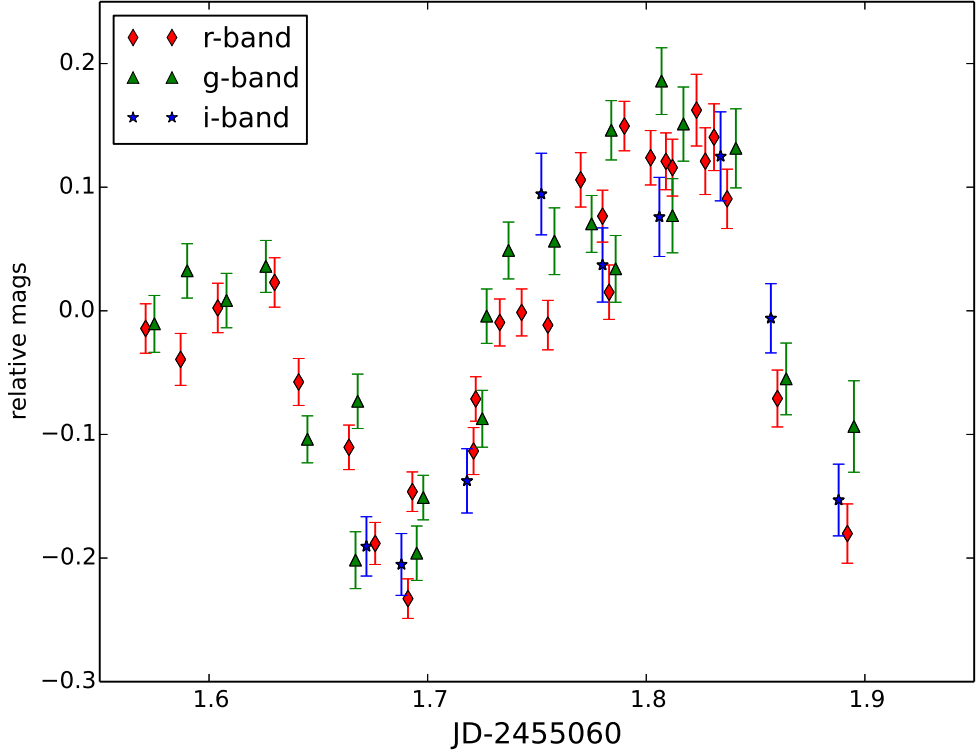}
\end{center}
\caption{Preliminary lightcurve from 18 and 19 August 2009 Palomar data. The
  magnitudes are relative to 8 field stars (with the mean removed). r-band (red
  diamond) and g-band (green triangle) photometry was obtained on both nights,
  while the i-band (blue star) was acquired only on the first night. The r- and
  g-band magnitudes have been arbitrarily adjusted to the same mean to show
  that there is no strong rotational colour dependence. The amplitude is
  $\sim$0.4 mag. Observations acquired on the 19 August 2009 have been
  arbitrarily shifted by 26 hours. This plot shows that the period is around
  6.5 hour if single peaked or around if double peaked 13 hour. Although the
  single peaked solutino seems incompatible with this plot, the quality of the
  data does not allow to reject it firmly. Thus one needs a longer time span to
  really characterize the lightcurve. }
\label{fig:ms9-3}
\end{figure*}

At its discovery distance of 13~AU, no coma has been detected in analysis of
our deep August 2009 CFHT images, to a limit of 28~mag/arcsec$^2$.  Other
shorter-period comets have been observed to start cometary activity as far out
as 12--14~AU from the Sun \citep[1P/Halley at 14~AU and 2060 Chiron at
  12~AU;][]{2004come.book..317M}. We observed (418993) at the Palomar 5m in
August 2009, and determined that it has a $\sim0.4$-mag lightcurve with a
period of over either 6.5 (single peaked) or 13 hours (double peaked;
Fig.~\ref{fig:ms9-3}). Studying a possible cometary activity on this object
requires determining the rotational phase to remove this predictable brightness
change. We obtained snapshot observations to monitor the cometary activity from
Aug. 2010 to Feb 2011 but detected none. From 2012 until end of 2014, many
observations of (418993) have been reported to MPC, around its perihelion
passage, but none have reported detection of cometary activity.

\afterpage{\clearpage}

\section{Summary and discussion}

The HiLat survey was designed to address one of the shortcomings of CFEPS, its
lack of sensitivity to high-inclination objects. HiLat imaged about
700~sqr.~deg. from 12$^\circ$ to 85$^\circ$ ecliptic latitude. The survey was
performed at CFHT in the $\mpr$ filter and achieved limiting magnitudes raging
from $\mpr = 22.4$ for the shallowest field to $\mpr = 24.8$ for the deepest
field. Being at high ecliptic latitude, the survey detected only {\lKBO}
objects, of which {\NKBO} are brighter than the characterization limit. Thanks
to the small number of objects and to our careful follow-up strategy, we
tracked all characterized objects to precise orbit determination and orbital
classification.

HiLat detected 6 objects from the hot main classical belt. We confirm the
global parameterization of this component found by CFEPS. An important finding
of CFEPS was that the $q$-distribution of the hot classical component is
essentially flat between 35~AU and 40~AU, with poor constraint on
this upper limit. The HiLat sample requires us to move the upper limit to
41~AU. Including the HiLat sample and survey in the analysis, we decrease
sightly the width of the inclination distribution of the hot component to
$\sigma = 14.5^\circ$.

The high sensitivity of HiLat survey to TNOs on highly-inclined orbits permits
formal rejection at high confidence of 'wider' orbital $i$-distributions for
the hot classical belt, and to a lesser extent the detached components. CFEPS
survey already rejected 'narrower' $i$ distributions. Having an
$i$-distribution with little contribution below about 10$^\circ$ and not
extending much beyond 35$^\circ$--40$^\circ$ is difficult to achieve with a
broad gaussian centered at $0^\circ$ distribution. It becomes increasingly
clear that eq.~(7) in \citet{2001AJ....121.2804B} is not the approrpiate
representation for this distribution and something different should be
considered. The distribution proposed by \citet{2010AJ....140..350G} is an
interesting possibility. A new $i$-distribution could have profound cosmogonic
implications that would need to be investigated.

The exotic higher-$i$ objects like those found in HiLat (Fig.~\ref{fig:ms9-1})
do not fit into this picture; we will call these $i\sim90^\circ$ objects the
`halo' component. Due to our sensitivity to high inclinations, these do not
represent the tail of the 14.5$^\circ$ gaussian.  Instead, these objects may
point to a new source that feeds large-$i$ TNOs into the planetary system
\citep{2009ApJ...697L..91G}. This may simultaneously be the source of the
Halley-Type comets \citep[see][]{2006Icar..184..619L}. Recently,
\citet{2016AJ....151...22B} pointed to (418993) as possible evidence that
this source might be related to an undiscovered planet in the 
distant solar system ($a\sim500$~au); producing $a<50$ objects like
2008~KV$_{42}$ requires pulling objects from such a large-$a$ source down to
such small semimajor axes and is exceedingly difficult due to the high
encounter speeds with Neptune and Uranus \citep{2009ApJ...697L..91G}.

The OSSOS Survey \citep{2016AJ.....inpressB,2016inprep........B} will allow a
careful consideration of the details of the $i$-distribution of the main hot
component and the relative fraction of objects that must be in this halo
population.  The use of our characterized Hilat survey (coupled to CFEPS and
OSSOS) permits powerful constraints to be placed on the $a/q/i$ distribution
generated by any proposed model of where these extreme objects are coming from.



\appendix
\section{Appendix A}
\label{sec:app_a}

We here detail the minor tuning to the L7 algorithm used to generate the hot
population of the main classical belt, motivated by the HiLat sample's greater
sensitvity.
The new algorithm becomes:
\begin{itemize}
\item a perihelion distance $q$ distribution that is mostly uniform between
  35 and 41~AU, with soft shoulders at both ends extending over $\sim$1~AU; the
  PDF is proportional to 
   $1/([1+\exp{((35 - q)/0.5)}] [1+\exp{((q - 41)/0.5)}])$; 
  any object with $q<$35~AU is rejected;
\item reject objects with $q < 38 - 0.2 i$~(deg) to account for 
  weaker long-term stability of low-$q$ orbits at low inclination.
\end{itemize}
The inclination distribution for the hot component remains $P(i) \propto
\sin(i) \exp{({-i^2}/{2 \sigma^2})}$, but with $\sigma = 14.5^\circ$.

\acknowledgements 
\noindent
{\bf Acknowledgments}: 
This work is based on observations obtained with MegaPrime/MegaCam, a joint
project of CFHT and CEA/DAPNIA, at the Canada-France-Hawaii Telescope (CFHT)
which is operated by the National Research Council (NRC) of Canada, the
Institute National des Sciences de l'Universe of the Centre National de la
Recherche Scientifique (CNRS) of France, and the University of Hawaii.
This research was supported by funding from the Natural Sciences and
Engineering Research Council of Canada, the Canadian Foundation for Innovation,
the National Research Council of Canada, and NASA Planetary Astronomy Program
NNG04GI29G.
This project could not have been a success without the dedicated staff of the
Canada-France-Hawaii telescope as well as the assistance of the skilled
telescope operators at KPNO and Mount Palomar.
This work is based in part on data produced and hosted at the Canadian
Astronomy Data Centre.

{\it Facilities:} \facility{CFHT (MegaPrime)}, \facility{WIYN},
\facility{Hale}, \facility{KPNO:2.1m}, \facility{Blanco}, \facility{MMT},
\facility{Gemini:South}, \facility{Subaru}, \facility{Magellan:Clay},
\facility{Gemini:Gillett (GMOS)}, \facility{NOT}

\eject
\newpage
\bibliographystyle{apj}
\bibliography{OtherPublications,Petit,Petit-MPEC}

\begin{thebibliography}{54}
\expandafter\ifx\csname natexlab\endcsname\relax\def\natexlab#1{#1}\fi

\bibitem[{{Allen} {et~al.}(2006){Allen}, {Gladman}, {Kavelaars}, {Petit},
  {Parker}, \& {Nicholson}}]{2006ApJ...640L..83A}
{Allen}, R.~L., {Gladman}, B., {Kavelaars}, J.~J., {Petit}, J., {Parker},
  J.~W., \& {Nicholson}, P. 2006, \apjl, 640, L83

\bibitem[{{Bannister} {et~al.}(2016{\natexlab{a}}){Bannister}, {Kavelaars},
  {Petit}, {Gladman}, {Gwyn}, {Chen}, {Volk}, {Alexandersen}, {Benecchi},
  {Delsanti}, {Fraser}, {Granvik}, {Grundy}, {Guilbert-Lepoutre}, {Hestroffer},
  {Ip}, {Jakubik}, {Jones}, {Kaib}, {Lacerda}, {Lawler}, {Lehner}, {Lin},
  {Lister}, {Lykawka}, {Monty}, {Marsset}, {Murray-Clay}, {Noll}, {Parker},
  {Pike}, {Rousselot}, {Rusk}, {Schwamb}, {Shankman}, {Sicardy}, {Vernazza}, \&
  {Wang}}]{2016AJ.....inpressB}
{Bannister}, M.~T., {Kavelaars}, J.~J., {Petit}, J.-M., {Gladman}, B.~J.,
  {Gwyn}, S.~D.~J., {Chen}, Y.-T., {Volk}, K., {Alexandersen}, M., {Benecchi},
  S., {Delsanti}, A., {Fraser}, W., {Granvik}, M., {Grundy}, W.~M.,
  {Guilbert-Lepoutre}, A., {Hestroffer}, D., {Ip}, W.-H., {Jakubik}, M.,
  {Jones}, L., {Kaib}, N., {Lacerda}, P., {Lawler}, S., {Lehner}, M.~J., {Lin},
  H.~W., {Lister}, T., {Lykawka}, P.~S., {Monty}, S., {Marsset}, M.,
  {Murray-Clay}, R., {Noll}, K., {Parker}, A., {Pike}, R.~E., {Rousselot}, P.,
  {Rusk}, D., {Schwamb}, M.~E., {Shankman}, C., {Sicardy}, B., {Vernazza}, P.,
  \& {Wang}, S.-Y. 2016{\natexlab{a}}, \aj

\bibitem[{{Bannister} {et~al.}(2016{\natexlab{b}}){Bannister}, {Kavelaars},
  {Petit}, {Gladman}, {Gwyn}, {Chen}, {Volk}, {Alexandersen}, {Benecchi},
  {Delsanti}, {Fraser}, {Granvik}, {Grundy}, {Guilbert-Lepoutre}, {Hestroffer},
  {Ip}, {Jakubik}, {Jones}, {Kaib}, {Lacerda}, {Lawler}, {Lehner}, {Lin},
  {Lister}, {Lykawka}, {Monty}, {Marsset}, {Murray-Clay}, {Noll}, {Parker},
  {Pike}, {Rousselot}, {Rusk}, {Schwamb}, {Shankman}, {Sicardy}, {Vernazza}, \&
  {Wang}}]{2016inprep........B}
---. 2016{\natexlab{b}}, In preparation

\bibitem[{{Batygin} \& {Brown}(2016)}]{2016AJ....151...22B}
{Batygin}, K. \& {Brown}, M.~E. 2016, \aj, 151, 22

\bibitem[{{Becker} {et~al.}(2008){Becker}, {Arraki}, {Kaib}, {Wood-Vasey},
  {Aguilera}, {Blackman}, {Blondin}, {Challis}, {Clocchiatti}, {Covarrubias},
  {Damke}, {Davis}, {Filippenko}, {Foley}, {Garg}, {Garnavich}, {Hicken},
  {Jha}, {Kirshner}, {Krisciunas}, {Leibundgut}, {Li}, {Matheson}, {Miceli},
  {Miknaitis}, {Narayan}, {Pignata}, {Prieto}, {Rest}, {Riess}, {Salvo},
  {Schmidt}, {Smith}, {Sollerman}, {Spyromilio}, {Stubbs}, {Suntzeff}, {Tonry},
  \& {Zenteno}}]{2008ApJ...682L..53B}
{Becker}, A.~C., {Arraki}, K., {Kaib}, N.~A., {Wood-Vasey}, W.~M., {Aguilera},
  C., {Blackman}, J.~W., {Blondin}, S., {Challis}, P., {Clocchiatti}, A.,
  {Covarrubias}, R., {Damke}, G., {Davis}, T.~M., {Filippenko}, A.~V., {Foley},
  R.~J., {Garg}, A., {Garnavich}, P.~M., {Hicken}, M., {Jha}, S., {Kirshner},
  R.~P., {Krisciunas}, K., {Leibundgut}, B., {Li}, W., {Matheson}, T.,
  {Miceli}, A., {Miknaitis}, G., {Narayan}, G., {Pignata}, G., {Prieto}, J.~L.,
  {Rest}, A., {Riess}, A.~G., {Salvo}, M.~E., {Schmidt}, B.~P., {Smith}, R.~C.,
  {Sollerman}, J., {Spyromilio}, J., {Stubbs}, C.~W., {Suntzeff}, N.~B.,
  {Tonry}, J.~L., \& {Zenteno}, A. 2008, \apjl, 682, L53

\bibitem[{{Bernstein} \& {Khushalani}(2000)}]{2000AJ....120.3323B}
{Bernstein}, G. \& {Khushalani}, B. 2000, \aj, 120, 3323

\bibitem[{{Bernstein} {et~al.}(2004){Bernstein}, {Trilling}, {Allen}, {Brown},
  {Holman}, \& {Malhotra}}]{2004AJ....128.1364B}
{Bernstein}, G.~M., {Trilling}, D.~E., {Allen}, R.~L., {Brown}, M.~E.,
  {Holman}, M., \& {Malhotra}, R. 2004, \aj, 128, 1364

\bibitem[{{Brasser} {et~al.}(2012){Brasser}, {Duncan}, {Levison}, {Schwamb}, \&
  {Brown}}]{2012Icar..217....1B}
{Brasser}, R., {Duncan}, M.~J., {Levison}, H.~F., {Schwamb}, M.~E., \& {Brown},
  M.~E. 2012, \icarus, 217, 1

\bibitem[{{Brown}(2001)}]{2001AJ....121.2804B}
{Brown}, M.~E. 2001, \aj, 121, 2804

\bibitem[{{Brown} {et~al.}(2005){Brown}, {Trujillo}, \&
  {Rabinowitz}}]{2005ApJ...635L..97B}
{Brown}, M.~E., {Trujillo}, C.~A., \& {Rabinowitz}, D.~L. 2005, \apjl, 635, L97

\bibitem[{{Doressoundiram} {et~al.}(2002){Doressoundiram}, {Peixinho}, {de
  Bergh}, {Fornasier}, {Th{\'e}bault}, {Barucci}, \&
  {Veillet}}]{2002AJ....124.2279D}
{Doressoundiram}, A., {Peixinho}, N., {de Bergh}, C., {Fornasier}, S.,
  {Th{\'e}bault}, P., {Barucci}, M.~A., \& {Veillet}, C. 2002, \aj, 124, 2279

\bibitem[{{Fraser} \& {Brown}(2012)}]{2012ApJ...749...33F}
{Fraser}, W.~C. \& {Brown}, M.~E. 2012, \apj, 749, 33

\bibitem[{{Fraser} {et~al.}(2014){Fraser}, {Brown}, {Morbidelli}, {Parker}, \&
  {Batygin}}]{2014ApJ...782..100F}
{Fraser}, W.~C., {Brown}, M.~E., {Morbidelli}, A., {Parker}, A., \& {Batygin},
  K. 2014, \apj, 782, 100

\bibitem[{{Fukugita} {et~al.}(1996){Fukugita}, {Ichikawa}, {Gunn}, {Doi},
  {Shimasaku}, \& {Schneider}}]{1996AJ....111.1748F}
{Fukugita}, M., {Ichikawa}, T., {Gunn}, J.~E., {Doi}, M., {Shimasaku}, K., \&
  {Schneider}, D.~P. 1996, \aj, 111, 1748

\bibitem[{{Gladman} \& {Chan}(2006)}]{2006ApJ...643L.135G}
{Gladman}, B. \& {Chan}, C. 2006, \apjl, 643, L135

\bibitem[{{Gladman} {et~al.}(2009){Gladman}, {Kavelaars}, {Petit}, {Ashby},
  {Parker}, {Coffey}, {Jones}, {Rousselot}, \& {Mousis}}]{2009ApJ...697L..91G}
{Gladman}, B., {Kavelaars}, J., {Petit}, J., {Ashby}, M.~L.~N., {Parker}, J.,
  {Coffey}, J., {Jones}, R.~L., {Rousselot}, P., \& {Mousis}, O. 2009, \apjl,
  697, L91

\bibitem[{{Gladman} {et~al.}(1998){Gladman}, {Kavelaars}, {Nicholson},
  {Loredo}, \& {Burns}}]{1998AJ....116.2042G}
{Gladman}, B., {Kavelaars}, J.~J., {Nicholson}, P.~D., {Loredo}, T.~J., \&
  {Burns}, J.~A. 1998, \aj, 116, 2042

\bibitem[{{Gladman} {et~al.}(2001){Gladman}, {Kavelaars}, {Petit},
  {Morbidelli}, {Holman}, \& {Loredo}}]{2001AJ....122.1051G}
{Gladman}, B., {Kavelaars}, J.~J., {Petit}, J.-M., {Morbidelli}, A., {Holman},
  M.~J., \& {Loredo}, T. 2001, \aj, 122, 1051

\bibitem[{{Gladman} {et~al.}(2012){Gladman}, {Lawler}, {Petit}, {Kavelaars},
  {Jones}, {Parker}, {Van Laerhoven}, {Nicholson}, {Rousselot}, {Bieryla}, \&
  {Ashby}}]{2012AJ....144...23G}
{Gladman}, B., {Lawler}, S.~M., {Petit}, J.-M., {Kavelaars}, J., {Jones},
  R.~L., {Parker}, J.~W., {Van Laerhoven}, C., {Nicholson}, P., {Rousselot},
  P., {Bieryla}, A., \& {Ashby}, M.~L.~N. 2012, \aj, 144, 23

\bibitem[{{Gladman} {et~al.}(2008){Gladman}, {Marsden}, \& {van
  Laerhoven}}]{2008ssbn.book...43G}
{Gladman}, B.~J., {Marsden}, B.~G., \& {van Laerhoven}, C. 2008, in The Solar
  System Beyond Neptune, ed. A.~{Barucci}, H.~{Boehnhardt}, D.~{Cruikshank}, \&
  A.~{Morbidelli}, LPI (Tucson: University of Arizona Press), 43--57

\bibitem[{{Gulbis} {et~al.}(2010){Gulbis}, {Elliot}, {Adams}, {Benecchi},
  {Buie}, {Trilling}, \& {Wasserman}}]{2010AJ....140..350G}
{Gulbis}, A.~A.~S., {Elliot}, J.~L., {Adams}, E.~R., {Benecchi}, S.~D., {Buie},
  M.~W., {Trilling}, D.~E., \& {Wasserman}, L.~H. 2010, \aj, 140, 350

\bibitem[{{Ida} {et~al.}(2000){Ida}, {Larwood}, \&
  {Burkert}}]{2000ApJ...528..351I}
{Ida}, S., {Larwood}, J., \& {Burkert}, A. 2000, \apj, 528, 351

\bibitem[{{Jewitt} {et~al.}(1996){Jewitt}, {Luu}, \&
  {Chen}}]{1996AJ....112.1225J}
{Jewitt}, D., {Luu}, J., \& {Chen}, J. 1996, \aj, 112, 1225

\bibitem[{{Jewitt} {et~al.}(1998){Jewitt}, {Luu}, \&
  {Trujillo}}]{1998AJ....115.2125J}
{Jewitt}, D., {Luu}, J., \& {Trujillo}, C. 1998, \aj, 115, 2125

\bibitem[{{Jones} {et~al.}(2006){Jones}, {Gladman}, {Petit}, {Rousselot},
  {Mousis}, {Kavelaars}, {Campo Bagatin}, {Bernabeu}, {Benavidez}, {Parker},
  {Nicholson}, {Holman}, {Grav}, {Doressoundiram}, {Veillet}, {Scholl}, \&
  {Mars}}]{2006Icar..185..508J}
{Jones}, R.~L., {Gladman}, B., {Petit}, J., {Rousselot}, P., {Mousis}, O.,
  {Kavelaars}, J.~J., {Campo Bagatin}, A., {Bernabeu}, G., {Benavidez}, P.,
  {Parker}, J.~W., {Nicholson}, P., {Holman}, M., {Grav}, T., {Doressoundiram},
  A., {Veillet}, C., {Scholl}, H., \& {Mars}, G. 2006, Icarus, 185, 508

\bibitem[{{Jones} {et~al.}(2010){Jones}, {Parker}, {Bieryla}, {Marsden},
  {Gladman}, {Kavelaars}, \& {Petit}}]{2010AJ....139.2249J}
{Jones}, R.~L., {Parker}, J.~W., {Bieryla}, A., {Marsden}, B.~G., {Gladman},
  B., {Kavelaars}, J., \& {Petit}, J. 2010, \aj, 139, 2249

\bibitem[{{Kaib} {et~al.}(2011){Kaib}, {Ro{\v s}kar}, \&
  {Quinn}}]{2011Icar..215..491K}
{Kaib}, N.~A., {Ro{\v s}kar}, R., \& {Quinn}, T. 2011, \icarus, 215, 491

\bibitem[{{Kavelaars} {et~al.}(2008){Kavelaars}, {Jones}, {Gladman}, {Parker},
  \& {Petit}}]{2008ssbn.book...59K}
{Kavelaars}, J., {Jones}, L., {Gladman}, B., {Parker}, J.~W., \& {Petit}, J.
  {The Orbital and Spatial Distribution of the Kuiper Belt}, ed. {Barucci,
  M.~A., Boehnhardt, H., Cruikshank, D.~P., \& Morbidelli, A. }, 59--69

\bibitem[{{Kavelaars} {et~al.}(2009){Kavelaars}, {Jones}, {Gladman}, {Petit},
  {Parker}, {Van Laerhoven}, {Nicholson}, {Rousselot}, {Scholl}, {Mousis},
  {Marsden}, {Benavidez}, {Bieryla}, {Campo Bagatin}, {Doressoundiram},
  {Margot}, {Murray}, \& {Veillet}}]{2009AJ....137.4917K}
{Kavelaars}, J.~J., {Jones}, R.~L., {Gladman}, B.~J., {Petit}, J., {Parker},
  J.~W., {Van Laerhoven}, C., {Nicholson}, P., {Rousselot}, P., {Scholl}, H.,
  {Mousis}, O., {Marsden}, B., {Benavidez}, P., {Bieryla}, A., {Campo Bagatin},
  A., {Doressoundiram}, A., {Margot}, J.~L., {Murray}, I., \& {Veillet}, C.
  2009, \aj, 137, 4917

\bibitem[{{Kenyon} \& {Bromley}(2004)}]{2004Natur.432..598K}
{Kenyon}, S.~J. \& {Bromley}, B.~C. 2004, \nat, 432, 598

\bibitem[{{Lamy} {et~al.}(2004){Lamy}, {Jorda}, {Toth}, {Weaver}, {Cruikshank},
  \& {Fernandez}}]{2004cosp...35.1824L}
{Lamy}, P.~L., {Jorda}, L., {Toth}, I., {Weaver}, H.~A., {Cruikshank}, D., \&
  {Fernandez}, Y. 2004, in COSPAR Meeting, Vol.~35, 35th COSPAR Scientific
  Assembly, ed. J.-P. {Paill{\'e}}, 1824

\bibitem[{{Levison} {et~al.}(2010){Levison}, {Duncan}, {Brasser}, \&
  {Kaufmann}}]{2010Sci...329..187L}
{Levison}, H.~F., {Duncan}, M.~J., {Brasser}, R., \& {Kaufmann}, D.~E. 2010,
  Science, 329, 187

\bibitem[{{Levison} {et~al.}(2006){Levison}, {Duncan}, {Dones}, \&
  {Gladman}}]{2006Icar..184..619L}
{Levison}, H.~F., {Duncan}, M.~J., {Dones}, L., \& {Gladman}, B.~J. 2006,
  \icarus, 184, 619

\bibitem[{{Levison} {et~al.}(2008){Levison}, {Morbidelli}, {Vanlaerhoven},
  {Gomes}, \& {Tsiganis}}]{2008Icar..196..258L}
{Levison}, H.~F., {Morbidelli}, A., {Vanlaerhoven}, C., {Gomes}, R., \&
  {Tsiganis}, K. 2008, Icarus, 196, 258

\bibitem[{{Levison} \& {Stern}(2001)}]{2001AJ....121.1730L}
{Levison}, H.~F. \& {Stern}, S.~A. 2001, \aj, 121, 1730

\bibitem[{{Magnier} \& {Cuillandre}(2004)}]{2004PASP..116..449M}
{Magnier}, E.~A. \& {Cuillandre}, J.-C. 2004, \pasp, 116, 449

\bibitem[{{Meech} \& {Svoren}(2004)}]{2004come.book..317M}
{Meech}, K.~J. \& {Svoren}, J. {Using cometary activity to trace the physical
  and chemical evolution of cometary nuclei}, ed. M.~C. {Festou}, H.~U.
  {Keller}, \& H.~A. {Weaver}, 317--335

\bibitem[{{Millis} {et~al.}(2002){Millis}, {Buie}, {Wasserman}, {Elliot},
  {Kern}, \& {Wagner}}]{2002AJ....123.2083M}
{Millis}, R.~L., {Buie}, M.~W., {Wasserman}, L.~H., {Elliot}, J.~L., {Kern},
  S.~D., \& {Wagner}, R.~M. 2002, \aj, 123, 2083

\bibitem[{{Morbidelli} \& {Levison}(2004)}]{2004AJ....128.2564M}
{Morbidelli}, A. \& {Levison}, H.~F. 2004, \aj, 128, 2564

\bibitem[{{Nesvorny}(2015)}]{2015AJ....150...73N}
{Nesvorny}, D. 2015, \aj, 150, 73

\bibitem[{{Peixinho} {et~al.}(2015){Peixinho}, {Delsanti}, \&
  {Doressoundiram}}]{2015A&A...577A..35P}
{Peixinho}, N., {Delsanti}, A., \& {Doressoundiram}, A. 2015, \aap, 577, A35

\bibitem[{{Petit} {et~al.}(2004){Petit}, {Holman}, {Scholl}, {Kavelaars}, \&
  {Gladman}}]{2004MNRAS.347..471P}
{Petit}, J., {Holman}, M., {Scholl}, H., {Kavelaars}, J., \& {Gladman}, B.
  2004, \mnras, 347, 471

\bibitem[{{Petit} {et~al.}(2006){Petit}, {Holman}, {Gladman}, {Kavelaars},
  {Scholl}, \& {Loredo}}]{2006MNRAS.365..429P}
{Petit}, J., {Holman}, M.~J., {Gladman}, B.~J., {Kavelaars}, J.~J., {Scholl},
  H., \& {Loredo}, T.~J. 2006, \mnras, 365, 429

\bibitem[{{Petit} {et~al.}(2008){Petit}, {Kavelaars}, {Gladman}, \&
  {Loredo}}]{2008ssbn.book...71P}
{Petit}, J., {Kavelaars}, J.~J., {Gladman}, B., \& {Loredo}, T. {Size
  Distribution of Multikilometer Transneptunian Objects}, ed. {Barucci, M.~A.,
  Boehnhardt, H., Cruikshank, D.~P., \& Morbidelli, A. }, 71--87

\bibitem[{{Petit} {et~al.}(2015){Petit}, {Allen}, {Gladman}, {Kavelaars},
  {Nicholson}, {Jacobson}, {Brozovic}, {Lawler}, {Parker}, \&
  {Williams}}]{2015MPEC....D....1P}
{Petit}, J.-M., {Allen}, L., {Gladman}, B., {Kavelaars}, J., {Nicholson}, P.,
  {Jacobson}, R., {Brozovic}, M., {Lawler}, S., {Parker}, J.~W., \& {Williams},
  G.~V. 2015, Minor Planet Electronic Circulars, 1

\bibitem[{{Petit} {et~al.}(2011){Petit}, {Kavelaars}, {Gladman}, {Jones},
  {Parker}, {Van Laerhoven}, {Nicholson}, {Mars}, {Rousselot}, {Mousis},
  {Marsden}, {Bieryla}, {Taylor}, {Ashby}, {Benavidez}, {Campo Bagatin}, \&
  {Bernabeu}}]{2011AJ....142..131P}
{Petit}, J.-M., {Kavelaars}, J.~J., {Gladman}, B.~J., {Jones}, R.~L., {Parker},
  J.~W., {Van Laerhoven}, C., {Nicholson}, P., {Mars}, G., {Rousselot}, P.,
  {Mousis}, O., {Marsden}, B., {Bieryla}, A., {Taylor}, M., {Ashby}, M.~L.~N.,
  {Benavidez}, P., {Campo Bagatin}, A., \& {Bernabeu}, G. 2011, \aj, 142, 131

\bibitem[{{Pike} {et~al.}(2015){Pike}, {Kavelaars}, {Petit}, {Gladman},
  {Alexandersen}, {Volk}, \& {Shankman}}]{2015AJ....149..202P}
{Pike}, R.~E., {Kavelaars}, J.~J., {Petit}, J.~M., {Gladman}, B.~J.,
  {Alexandersen}, M., {Volk}, K., \& {Shankman}, C.~J. 2015, \aj, 149, 202

\bibitem[{{Press} {et~al.}(1992){Press}, {Teukolsky}, {Vetterling}, \&
  {Flannery}}]{1992nrfa.book.....P}
{Press}, W.~H., {Teukolsky}, S.~A., {Vetterling}, W.~T., \& {Flannery}, B.~P.
  1992, {Numerical recipes in FORTRAN. The art of scientific computing}

\bibitem[{{Tegler} {et~al.}(2003){Tegler}, {Romanishin}, \&
  {Consolmagno}}]{2003ApJ...599L..49T}
{Tegler}, S.~C., {Romanishin}, W., \& {Consolmagno}, G.~J. 2003, \apjl, 599,
  L49

\bibitem[{{Thommes} {et~al.}(1999){Thommes}, {Duncan}, \&
  {Levison}}]{1999Natur.402..635T}
{Thommes}, E.~W., {Duncan}, M.~J., \& {Levison}, H.~F. 1999, \nat, 402, 635

\bibitem[{{Trujillo} \& {Jewitt}(1998)}]{1998AJ....115.1680T}
{Trujillo}, C. \& {Jewitt}, D. 1998, \aj, 115, 1680

\bibitem[{{Trujillo} \& {Brown}(2003)}]{2003EM&P...92...99T}
{Trujillo}, C.~A. \& {Brown}, M.~E. 2003, Earth Moon and Planets, 92, 99

\bibitem[{{Trujillo} {et~al.}(2000){Trujillo}, {Jewitt}, \&
  {Luu}}]{2000ApJ...529L.103T}
{Trujillo}, C.~A., {Jewitt}, D.~C., \& {Luu}, J.~X. 2000, \apjl, 529, L103

\bibitem[{{Trujillo} {et~al.}(2001){Trujillo}, {Jewitt}, \&
  {Luu}}]{2001AJ....122..457T}
---. 2001, \aj, 122, 457

\end{thebibliography}

\end{document}